\documentclass[amsmath, amssymb, aps, reprint, superscriptaddress, nofootinbib]{revtex4-2}

\usepackage{graphicx}
\usepackage{dcolumn}  

\usepackage{bm}
\usepackage{physics}
\usepackage[svgnames]{xcolor}
\usepackage[colorlinks=true, linkcolor=blue, citecolor=blue, urlcolor=blue]{hyperref}
\usepackage{amsthm}
\usepackage{enumitem}

\begin{document}

\title{Electron--Photon Spatial Entanglement in Coherent Cathodoluminescence}

\author{Tatsuro Yuge}
\email{yuge.tatsuro@shizuoka.ac.jp}
\affiliation{Department of Physics, Shizuoka University, Shizuoka 422-8529, Japan}

\author{Ryo Okamoto}
\affiliation{Department of Electronic Science and Engineering, Kyoto University, Kyotodaigakukatsura, Nishikyo-ku, Kyoto 615-8510, Japan}

\author{Takumi Sannomiya}
\affiliation{Department of Materials Science and Engineering, School of Materials and Chemical Technology, Institute of Science Tokyo, Kanagawa, 4259 Nagatsuta, Midori-ku 226-8503, Yokohama Japan}

\author{Keiichirou Akiba}
\email{akiba.keiichiro@qst.go.jp}
\affiliation{Takasaki Institute for Advanced Quantum Science, National Institutes for Quantum Science and Technology, 1233 Watanuki, Takasaki, Gunma 370-1292, Japan}

\begin{abstract}
Electron--photon quantum entanglement in an electron microscope paves the way for a new quantum platform, enabling the integration of quantum functionalities into electron microscopy and opening opportunities for quantum imaging and quantum sensing at the nanoscale. To realize such a platform, it is crucial to understand the degree and nature of electron--photon entanglement in cathodoluminescence (CL). However, its dependence on electron-beam properties, particularly transverse coherence, remains unclear. Here, we present a theoretical framework describing the quantum state of an electron--photon pair generated in coherent CL from a system with translational invariance in the plane perpendicular to the electron beam, as can be realized, for example, in transition radiation at a planar interface. By expressing the scattered state directly in terms of the luminescence spectrum, we evaluate the entanglement using both subsystem purity and an Einstein--Podolsky--Rosen-type criterion. These two measures enable a clear distinction among wave-like, particle-like, and classical regimes in terms of spatial and momentum entanglement in the electron--photon system. Our analysis identifies the roles of the electron's transverse and longitudinal coherence, as well as the photon's spectral width, and reveals the conditions under which strong spatial entanglement emerges. This unified perspective clarifies the nature of electron--photon quantum correlations in coherent CL and advances the development of quantum-enabled functionalities in electron microscopy.
\end{abstract}

\maketitle

\section{Introduction}

Exploiting the wave nature of electrons, electron microscopy provides nanoscale visualization of materials combined with spectroscopic access to local material information such as elemental composition and electronic structure. The decisive experimental demonstration of the Aharonov-Bohm effect was enabled by the phase coherence of electron waves \cite{Tonomura_etal1986}. The Young's double-slit experiment in electron microscopy successfully captured the single-electron buildup process of the interference pattern, excellently demonstrating the wave-particle duality of a single electron \cite{Tonomura_etal1989}. Thus, quantum features in electron microscopy are fundamentally rooted in the wave nature of electrons.

Recent advances have brought about novel quantum aspects of swift electrons, necessitating the frameworks of quantum electrodynamics and quantum optics \cite{Ruimy_etal2025,Abajo_etal2025}. In such a framework, unprecedented technologies have been cultivated so far, including quantum-enhanced sensing and metrology beyond classical limits \cite{Giovannetti_etal2004,Giovannetti_etal2011,Degen_etal2017,Barbieri_2022}, as well as quantum computation and information processing \cite{NielsenChuang_2010,YamamotoSemba_2016,Wilde_2017}. Extending these concepts to electron microscopy offers novel opportunities for advancing its capabilities.

Indeed, this field of free-electron quantum optics is emerging, and electron systems in electron microscopy can serve as a new quantum platform, where the coherent interaction between electrons and photons is crucial for their realization \cite{Mechel_etal2021,Ruimy_etal2025}. Such electron--photon interactions in the quantum regime can also be exploited to advance nanophotonics \cite{Roques-Carmes_etal2023,Abajo_etal2025}. Pioneering studies have demonstrated that free electrons can be induced to emit and absorb integer numbers of photons through their interaction with laser fields \cite{Barwick_etal2009}. Subsequently, the coherent quantum state manipulation of free-electron populations has been realized via strong light fields \cite{Feist_etal2015}. These experimental results advanced the quantum description of electron--photon interactions. Theoretical studies have widely investigated the quantum aspect of scattering and generation of entanglement between an electron and a photon \cite{Kfir2019,Kfir_etal2021,Karnieli_etal2021,PanGover_2021,Konecna_etal2022,Huang_etal2023,Kazakevich_etal2024,Shi_etal2024,Henke_etal2025a,Rembold_etal2025}. In experimental studies, correlated electron--photon pairs have been demonstrated in transmission electron microscopes integrated with photonic systems, via cavity-mediated interaction \cite{Feist_etal2022} and via interaction with a membrane \cite{Preimesberger_etal2025a}.
Only recently have demonstrations of electron--photon entanglement generated via cathodoluminescence (CL) been reported \cite{Preimesberger_etal2025,Henke_etal2025b}.

These results mark a milestone in leveraging the unique quantum properties in electron microscopy. The introduction of quantum imaging techniques holds great potential for overcoming the critical issue of electron-beam induced damage \cite{OkamotoNagatani2014,Kruit_etal2016}. It is also expected that nonclassical light, which has not been achieved through conventional optical means, can be generated; for instance, unprecedented high-NOON states at megahertz rates \cite{VelascoAbajo2025}. Furthermore, theoretical investigations have proposed a variety of intriguing concepts for quantum information processing and quantum sensing \cite{Ruimy_etal2025,Reinhardt_etal2021,Karnieli_etal2024}. Towards these advances, methods for achieving strong coupling between electrons and photons have been investigated \cite{Talebi2020,Xie_etal2025,Zhao2025}. However, the degree and nature of electron--photon entanglement in CL are not fully understood. It remains unclear how the entanglement depends on electron beam and optical emission properties, especially on the transverse coherence of the electron beam. Moreover, different entanglement measures may give qualitatively different assessments of the entangled state, and a systematic understanding of these differences is still lacking.

\begin{figure}[tb]
  \centering
  \includegraphics[width=0.9\linewidth]{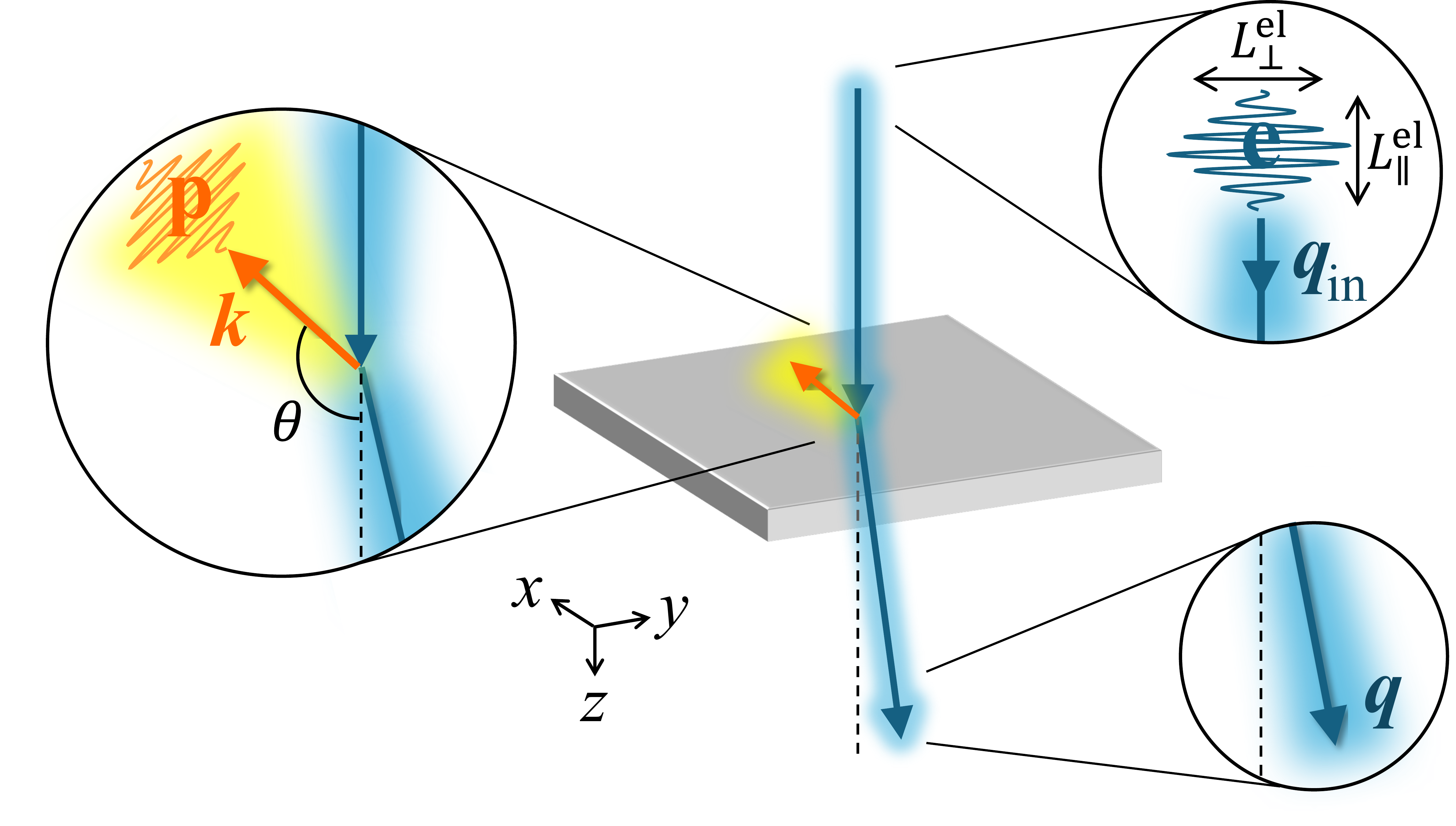}
  \caption{\label{fig:setup}
  Setup of coherent cathodoluminescence.
  The incident electron is described by a wave packet with momentum $\hbar \bm{q}_{\mathrm{in}}$, whose mean value is $\hbar q_0 \bm{e}_z$, and with widths $L_\perp^\mathrm{el}$ and $L_\parallel^\mathrm{el}$ in the transverse ($xy$) and longitudinal ($z$) directions, respectively.
  A radiated photon has a wavevector $\bm{k}$, centered around $k_c$ in magnitude (corresponding to the wavelength $\lambda_c = 2\pi / k_c$), with spectral width $\Delta \lambda_{\mathrm{ph}}$, and is emitted at an angle $\theta$.
  The electron recoils and has momentum $\hbar \bm{q}$ in the scattered state.
  }
\end{figure}

\begin{figure}[tb]
  \includegraphics[width=\linewidth]{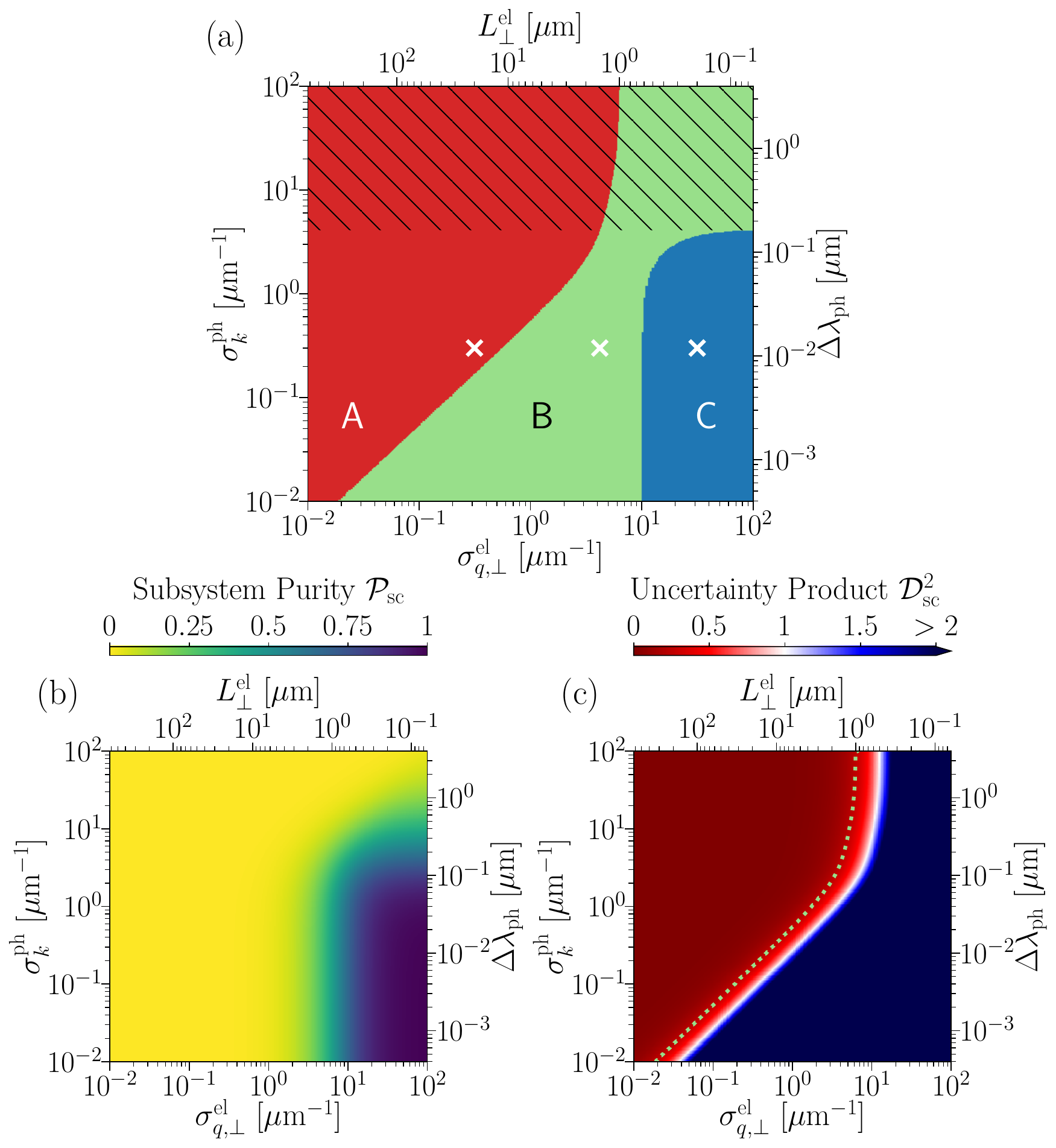}
  \caption{\label{fig:entanglement_1}
  (a) Parameter regimes of entanglement in the scattered state $\ket{\Psi_{\mathrm{sc}}}$ (Sec.~\ref{sec:scattered_state}) of the electron--photon system.
  Regimes A (red), B (light green), and C (blue) are defined based on (b) subsystem purity $\mathcal{P}_{\mathrm{sc}}$ (Sec.~\ref{sec:purity}) and (c) uncertainty product $\mathcal{D}_{\mathrm{sc}}^2$ (Sec.~\ref{sec:EPR}).
  Regime A corresponds to the wave-like entangled regime, where $\mathcal{P}_{\mathrm{sc}} < 2/3$ and $\mathcal{D}_{\mathrm{sc}}^2 < 1/4$;
  Regime B to the particle-like entangled regime, where $\mathcal{P}_{\mathrm{sc}} < 2/3$ and $\mathcal{D}_{\mathrm{sc}}^2 > 1/4$;
  and Regime C to the classical regime, where $\mathcal{P}_{\mathrm{sc}} > 2/3$ and $\mathcal{D}_{\mathrm{sc}}^2 > 1/4$.
  The upper hatched region indicates the entangled regime identified by the electron $z$-subsystem purity $\mathcal{P}^\mathrm{el}_z < 2/3$,
  i.e., entanglement between the electron's $z$-degree of freedom and the remaining degrees of freedom (Sec.~\ref{sec:longitudinal}).
  Here, we use $2/3$ as a practical threshold for both $\mathcal{P}_{\mathrm{sc}}$ and $\mathcal{P}^{\mathrm{el}}_z$ for visualization and classification. For $\mathcal{P}_{\mathrm{sc}}$, this threshold distinguishes the strongly entangled regimes from the nearly separable regime.
  The qualitative features in (a) remain unchanged if this threshold is chosen closer to unity.
  The dotted curve in (c) represents the EPR-correlation threshold $\mathcal{D}_{\mathrm{sc}}^2 = 1/4$.
  The horizontal and vertical axes represent the transverse width of the electron wavenumber $\sigma_{q,\perp}^{\mathrm{el}}$ and the luminescence spectral width $\sigma_k^{\mathrm{ph}}$, respectively.
  The parameters are $K = 200~\mathrm{keV}$, $\lambda_c = 0.5~\mathrm{\mu m}$, and $L_\parallel^\mathrm{el} = 1.3~\mathrm{\mu m}$.
  }
\end{figure}

In this paper, we identify, within a simple model of coherent CL from a sample with a planar surface (schematically illustrated in Fig.~\ref{fig:setup}), the physical conditions under which distinct forms of electron--photon entanglement arise.
We show that the scattered state exhibits three distinct entanglement regimes, as shown in Fig.~\ref{fig:entanglement_1}(a): two of them (A and B) correspond to different types of entangled states, while the third (C) is a non-entangled regime.
We further demonstrate that these regimes correspond to the wave-like, particle-like, and classical regimes recently proposed in Ref.~\cite{Huang_etal2023} for electron-energy--photon-number entanglement, thereby extending these concepts to spatial and momentum entanglement.
By clarifying the conditions under which strong electron--photon entanglement can be realized, our results provide a unified perspective on quantum correlations in CL and offer practical guidelines for generating spatially entangled electron--photon pairs.

To achieve this, we develop a theoretical framework to reconstruct the scattered state $\ket{\Psi_{\mathrm{sc}}}$ of electron--photon pairs directly from the CL spectrum. 
This framework assumes that the sample has a planar surface and is translationally invariant in the plane perpendicular to the electron beam, as depicted in Fig.~\ref{fig:setup}. 
Such a geometry can be realized, for example, in transition radiation at a planar interface.
Within this approach, we quantify entanglement using the subsystem purity and the EPR (Einstein-Podolsky-Rosen) criterion [see Figs.~\ref{fig:entanglement_1}(b) and (c)].

The paper is organized as follows.
Section~\ref{sec:interaction} introduces a theoretical model for electron--photon interactions in coherent CL and outlines the assumptions underlying our framework.
Section~\ref{sec:scattered_state} presents an expression for the scattered electron--photon state $\ket{\Psi_{\mathrm{sc}}}$ reconstructed from the luminescence spectrum.
Section~\ref{sec:entanglement} provides a quantitative analysis of electron--photon entanglement using the two measures. We identify wave-like, particle-like, and classical regimes and clarify the roles of the transverse and longitudinal electron coherence, as well as the CL spectral width. We also present the joint probability distributions of the relative position and total wavevector in the electron--photon state.
Finally, Sec.~\ref{sec:conclusion} summarizes our results and provides their implications.

\section{Electron--photon interaction}
\label{sec:interaction}

We consider coherent CL generated by an electron beam interacting with a sample \cite{Abajo2010}. As illustrated in Fig.~1, an electron beam propagating along the $z$ direction impinges on a sample whose planar surface is lies in the $xy$ plane. An incident electron has a momentum $\hbar\bm{q}_{\mathrm{in}}$ whose mean value is $\hbar q_0\bm{e}_z$ ($\bm{e}_z$ is the unit vector in the $z$ direction), and whose momentum widths are $\hbar \sigma_{q,\perp}^{\mathrm{el}}$ and $\hbar \sigma_{q,\parallel}^{\mathrm{el}}$ in the transverse $(xy)$ and longitudinal $(z)$ directions, respectively. The interaction of the electron with the sample excites the electromagnetic field, resulting in the emission of a photon with wavevector $\bm{k}$. In this work, we focus on the weak-coupling regime, in which each incident electron emits at most one photon. Owing to momentum exchange during the emission process, the electron recoils to a final momentum $\hbar\bm{q}$, which is correlated with the momentum of the emitted photon.

Within this setup, we analyze the scattering process between a single electron in the beam and the radiation field.
The total Hamiltonian of the electron--photon system
$\hat{H} = \hat{H}_0 + \hat{H}_\mathrm{int}$
is composed of the free Hamiltonian $\hat{H}_0 = \hat{\bm{p}}^2 / (2m)
  + \sum_{\bm{k}} \hbar\omega_{\bm{k}} \hat{a}_{\bm{k}}^\dagger \hat{a}_{\bm{k}}$
and the electron--photon interaction $\hat{H}_\mathrm{int}$.
Here $\hat{\bm{p}}$ is the momentum operator of the electron, $m$ is the electron mass, and $\hat{a}_{\bm{k}}$ ($\hat{a}_{\bm{k}}^\dagger$) is the annihilation (creation) operator of a photon with a wavevector $\bm{k}$.
The photon frequency is given by $\omega_{\bm{k}} = c k$, where $k = |\bm{k}|$ and $c$ is the speed of light.

In this work, we assume that the electron--photon system is translationally invariant in the transverse ($xy$) plane perpendicular to the electron beam and rotationally symmetric around the beam axis.
In the context of CL, such a situation can be realized in transition radiation (Fig.~\ref{fig:setup}), where the transverse extent of the sample is sufficiently large.
The translational invariance in the transverse plane implies that the in-plane momentum of the electron--photon system is conserved in the scattering process.
Consequently, a matrix element of the interaction Hamiltonian $\hat{H}_\mathrm{int}$ takes the form
\begin{align}
  \bra{\bm{q}, \bm{k}} \hat{H}_\mathrm{int} \ket{\bm{q}', \mathrm{vac}}
  = h(\bm{k}, q_z - q'_z) \, \delta(\bm{q}_\perp + \bm{k}_\perp - \bm{q}'_\perp),
  \label{in-plane-momentum-conservation}
\end{align}
where $\bm{q} = (q_x, q_y, q_z)$ and $\bm{q}' = (q'_x, q'_y, q'_z)$ are the electron wavevectors,
and $\bm{q}_\perp = (q_x, q_y)$ and $\bm{q}'_\perp = (q'_x, q'_y)$ denote their transverse components.
Similarly, $\bm{k} = (k_x, k_y, k_z)$ and $\bm{k}_\perp = (k_x, k_y)$ represent the photon wavevector and its transverse components, respectively.
The function $h(\bm{k}, q_z - q'_z)$ represents the interaction strength.
The states appearing in Eq.~\eqref{in-plane-momentum-conservation} are defined as follows:
$\ket{\bm{q}, \mathrm{vac}} = \ket{\bm{q}}_{\mathrm{el}} \ket{\mathrm{vac}}_{\mathrm{ph}}$ and
$\ket{\bm{q}, \bm{k}} = \ket{\bm{q}}_{\mathrm{el}} \ket{\bm{k}}_{\mathrm{ph}}$,
where
$\ket{\bm{q}}_{\mathrm{el}}$ is the eigenstate of $\hat{\bm{p}}$ satisfying $\hat{\bm{p}} \ket{\bm{q}}_{\mathrm{el}} = \hbar \bm{q} \ket{\bm{q}}_{\mathrm{el}}$,
$\ket{\mathrm{vac}}_{\mathrm{ph}}$ is the photon vacuum state satisfying $\hat{a}_{\bm{k}} \ket{\mathrm{vac}}_{\mathrm{ph}} = 0$,
and $\ket{\bm{k}}_{\mathrm{ph}} = \hat{a}_{\bm{k}}^\dag \ket{\mathrm{vac}}_{\mathrm{ph}}$ is a single-photon state.
Furthermore, the rotational symmetry implies
that the interaction strength satisfies $h(k_\perp \cos\phi, k_\perp \sin\phi, k_z, q_z - q'_z)$ being invariant under rotations $0 \le \phi < 2\pi$.

\section{Scattered state}
\label{sec:scattered_state}

We now analyze the quantum state of the electron--photon system generated by the scattering process.
We consider an initial state consisting of a single-electron wave packet and the photon vacuum,
$\ket{\Psi_\mathrm{ini}} = \int d^3 q_{\mathrm{in}} \, \psi_\mathrm{ini}(\bm{q}_{\mathrm{in}}) \ket{\bm{q}_{\mathrm{in}}, \mathrm{vac}}$,
where $\psi_\mathrm{ini}(\bm{q}_{\mathrm{in}})$ denotes the wavevector-space wave function of the incoming electron.

In the weak-coupling regime considered here, a first-order perturbation analysis in $\hat{H}_\mathrm{int}$ \cite{AbajoGiulio2021} yields that the electron--photon state after the scattering can be written as 
$\ket{\Psi_\mathrm{fin}} = \sqrt{1 - a} \ket{\Psi_\mathrm{ini}} + \sqrt{a} \ket{\Psi_{\mathrm{sc}}}$ ($0 < a < 1$).
Here $\ket{\Psi_{\mathrm{sc}}}$ is the scattered state associated with single-photon generation
(see Appendix~\ref{appendix:perturbation} for the detailed derivation).

Furthermore, using the interaction matrix element given in Eq.~\eqref{in-plane-momentum-conservation},
we find that $\ket{\Psi_{\mathrm{sc}}}$ can be reconstructed from the excitation probability density $\Gamma(\bm{k})$ of a photon with $\bm{k}$ as (see Appendix~\ref{appendix:state_luminescence} for the detailed derivation)
\begin{align}
  \ket{\Psi_{\mathrm{sc}}}
   & = \int d^3q \, d^3k \, \psi_\mathrm{sc}(\bm{q}, \bm{k}) \ket{\bm{q}, \bm{k}},
  \label{Psi_scattered}
  \\
  \psi_\mathrm{sc}(\bm{q}, \bm{k})
   & = \psi_\mathrm{ini}(\bm{q}_\perp + \bm{k}_\perp, q_z + ck/v_z) \sqrt{\Gamma(\bm{k})} e^{i\eta(\bm{k})}.
  \label{psi_sc}
\end{align}
In the above equation, $\psi_\mathrm{ini}(\bm{q}_\perp + \bm{k}_\perp, q_z + ck/v_z)$ denotes $\psi_\mathrm{ini}(\bm{q}_{\mathrm{in}}) \big|_{\bm{q}_{\mathrm{in},\perp} = \bm{q}_\perp + \bm{k}_\perp, \, q_{\mathrm{in},z} = q_z + ck/v_z}$, where $\bm{q}_{\mathrm{in}} = (q_{\mathrm{in},x}, q_{\mathrm{in},y}, q_{\mathrm{in},z})$ and $\bm{q}_{\mathrm{in},\perp} = (q_{\mathrm{in},x}, q_{\mathrm{in},y})$.
Here, $v_z$ is the mean velocity of an incident electron and $\eta(\bm{k})$ is an undetermined phase that cannot be inferred from the luminescence spectrum $\Gamma(\bm{k})$ alone and that may depend on $\bm{k}$.
We note that the phase $\eta(\bm{k})$ can be eliminated by a local operation on the photon.

We also note that the scattered electron--photon state $\ket{\Psi_{\mathrm{sc}}}$ is pure. 
This is justified in the weak-coupling regime when photon-emission events are postselected, processes in which photon emission is accompanied by additional nonradiative excitations are negligible to the perturbative order considered, and the unobserved degrees of freedom do not retain information that distinguishes different electron--photon momentum components. 
In realistic samples, other inelastic processes, such as the excitation of surface plasmons, bulk plasmons, and phonons, may coexist with the radiative process considered here. Events involving these processes but no photon in the selected detection channel can be largely rejected by electron--photon coincidence detection. Such competing processes therefore primarily reduce the rate at which the desired radiative electron--photon pairs are successfully selected. A distinct limitation arises when photon emission is accompanied by unresolved nonradiative or environmental excitations. If these unobserved degrees of freedom retain information that distinguishes different electron--photon momentum components, the conditional electron--photon state generally becomes mixed.

This formulation can be straightforwardly extended to the case where a filter is applied to the photon.
Indeed, when a photonic filter with a transfer function $F(\bm{k})$ is applied to the scattered state $\ket{\Psi_{\mathrm{sc}}}$,
the filtered state, $\ket*{\tilde{\Psi}_{\mathrm{sc}}} = \sqrt{\mathcal{N}_F} \int d^3k \, F(\bm{k}) \, \ket{\bm{k}}_{\mathrm{ph}}\braket{\bm{k}}{\Psi_{\mathrm{sc}}}$ with $\mathcal{N}_F^{-1} = \int d^3k \, |F(\bm{k})|^2 \, \Gamma(\bm{k})$,
can be written as
$\ket*{\tilde{\Psi}_{\mathrm{sc}}} = \int d^3q \, d^3k \, \tilde{\psi}_\mathrm{sc}(\bm{q}, \bm{k}) \ket{\bm{q}, \bm{k}}$,
where
\begin{align}
   & \tilde{\psi}_\mathrm{sc}(\bm{q}, \bm{k})
  \\
   & = \psi_\mathrm{ini}(\bm{q}_\perp + \bm{k}_\perp, q_z + ck/v_z) F(\bm{k}) \sqrt{\mathcal{N}_F \Gamma(\bm{k})} e^{i\eta(\bm{k})}.
  \notag
\end{align}
By introducing the redefined quantities $\tilde{\Gamma}(\bm{k}) = \mathcal{N}_F |F(\bm{k})|^2 \Gamma(\bm{k})$ and $e^{i\tilde{\eta}(\bm{k})} = e^{i\eta(\bm{k})} F(\bm{k}) / |F(\bm{k})|$,
the filtered state $\ket*{\tilde{\Psi}_{\mathrm{sc}}}$ can be cast into the same form as Eqs.~\eqref{Psi_scattered} and \eqref{psi_sc}.
Hereafter, $\Gamma(\bm{k})$ and $\eta(\bm{k})$ are therefore understood to represent either the original or their filtered quantities, depending on the context.

In the following, we assume that the initial electron wave function $\psi_\mathrm{ini}(\bm{q}_{\mathrm{in}})$ has a Gaussian form:
$\psi_\mathrm{ini}(\bm{q}_{\mathrm{in}}) = \prod_{\alpha=x,y,z} \psi^{(\alpha)}_\mathrm{ini}(q_{\mathrm{in},\alpha})$, where
\begin{align}
  \psi^{(\alpha)}_\mathrm{ini}(q_{\mathrm{in},\alpha})
  = \frac{1}{(2\pi)^{1/4} (\sigma_{q,\perp}^\mathrm{el})^{1/2}} \exp[  -\frac{q_{\mathrm{in},\alpha}^2}{4 (\sigma_{q,\perp}^\mathrm{el})^2} ]
  \label{initial_psi_xy}
\end{align}
for $\alpha = x, y$, and
\begin{align}
  \psi^{(z)}_\mathrm{ini}(q_{\mathrm{in},z}) = \frac{1}{(2\pi)^{1/4} (\sigma_{q,\parallel}^\mathrm{el})^{1/2}} \exp[ -\frac{(q_{\mathrm{in},z} - q_0)^2}{4 (\sigma_{q,\parallel}^\mathrm{el})^2} ].
  \label{initial_psi_z}
\end{align}
This means that the electron has, on average, a wavevector $q_0 \bm{e}_z$,
with an uncertainty width of $\sigma_{q,\perp}^\mathrm{el}$ in the transverse ($xy$) direction
and $\sigma_{q,\parallel}^\mathrm{el}$ in the longitudinal ($z$) direction.
The central wavenumber $q_0$ is determined by the electron kinetic energy $K$ as $\hbar q_0 = (1/c)\sqrt{K^2 + 2mc^2 K}$ and is related to the mean velocity as $v_z = \hbar q_0/m$.
$\sigma_{q,\perp}^\mathrm{el}$ and $\sigma_{q,\parallel}^\mathrm{el}$ are estimated from the transverse and longitudinal coherence lengths of the electron beam as
$\sigma_{q,\perp}^\mathrm{el} = 2\pi / L_\perp^\mathrm{el}$ and $\sigma_{q,\parallel}^\mathrm{el} = 2\pi / L_\parallel^\mathrm{el}$, respectively.

The transverse coherence length of an electron beam depends on the electron source and illumination conditions. In conventional electron microscopy, $L_\perp^\mathrm{el}$ is typically in the nanometer to submicrometer range \cite{HatanakaYamasaki2021,HatanakaYamasaki2025}. Much larger values can, however, be achieved under highly collimated illumination. For example, in sensitivity-enhanced field-emission electron holography, an illuminating electron beam with an angular divergence of the order of $\alpha = 10^{-8}~\mathrm{rad}$ has been reported \cite{Hasegawa_etal1989}. From the estimate $L_\perp^\mathrm{el} \approx \lambda_\mathrm{el} / (2 \alpha)$ with the electron de Broglie wavelength $\lambda_\mathrm{el}$, 
this corresponds to a transverse coherence length on the order of several hundred micrometers for high-energy electrons.

Furthermore, for an illustrative calculation, we assume a simple phenomenological model for the luminescence spectrum $\Gamma(\bm{k})$.
Owing to the rotational symmetry in the transverse plane, $\Gamma(\bm{k})$ is independent of the azimuthal angle $\phi$.
We assume that the photon wavenumber $k$ of the luminescence is peaked around $k_c$ with a characteristic width $\sigma_{k}^\mathrm{ph}$. 
Furthermore, we assume an angular distribution in which photons are emitted only into the upper half-space and predominantly in oblique directions.
Specifically, we adopt the following form of $\Gamma(\bm{k})$ in the spherical coordinates:
\begin{align}
  \Gamma(\bm{k}) & = \Gamma(k, \theta, \phi) = g(k) f(\theta),
  \label{Gamma_model}
  \\
  g(k)           & = \mathcal{N}_g \exp[ -\frac{(k-k_c)^2}{2(\sigma_{k}^\mathrm{ph})^2}],
  \label{g_model}
  \\
  f(\theta)      & =
  \begin{cases}
    0 & 0 \le \theta \le \pi / 2
    \\[3pt]
    \displaystyle \frac{15}{4\pi} (\sin\theta \cos\theta)^2 & \pi / 2 < \theta \le \pi.
  \end{cases}
  \label{f_model}
\end{align}
The normalization constant $\mathcal{N}_g$ for $g(k)$ is defined to satisfy $\int_0^\infty dk \, k^2 g(k) = 1$,
so that
\begin{align}
  \mathcal{N}_g^{-1}
   & = \sqrt{\frac{\pi}{2}} \sigma_{k}^\mathrm{ph} \bigl[ (\sigma_{k}^\mathrm{ph})^2 + k_c^2 \bigr]
  \biggl[ \erf\biggl(\frac{k_c}{\sqrt{2} \sigma_{k}^\mathrm{ph}} \biggr) + 1 \biggr]
  \notag
  \\
   & \quad + k_c (\sigma_{k}^\mathrm{ph})^2 \exp\biggl( -\frac{k_c^2}{2 (\sigma_{k}^\mathrm{ph})^2}\biggr),
  \label{N_g}
\end{align}
where $\erf(\cdot)$ is the Gauss error function.
By this normalization, $\Gamma(\bm{k})$ is also normalized:
$\int d^3k \, \Gamma(\bm{k}) = \int dk \, d\theta \, d\phi \, k^2 \sin\theta \, \Gamma(k, \theta, \phi) = 1$.
We note that $k_c$ and $\sigma_{k}^\mathrm{ph}$ are related to the central wavelength $\lambda_c$ and the spectral width $\Delta \lambda_{\mathrm{ph}}$ via $k_c = 2 \pi / \lambda_c$ and $\sigma_{k}^\mathrm{ph} = 2 \pi \Delta \lambda_{\mathrm{ph}} / \lambda_c^2$, respectively.
When a photonic bandpass filter is applied, $\lambda_c$ and $\Delta \lambda_{\mathrm{ph}}$ correspond to the central wavelength and bandwidth of the filter, respectively.

We note that the luminescence spectrum $\Gamma(\bm{k})$ adopted in Eqs.~\eqref{Gamma_model}--\eqref{f_model} is a simplified model and is not intended to provide a quantitative description of transition radiation. 
In particular, unlike the Ginzburg--Frank formula \cite{GinzburgFrank1945,Ginzburg1979}, its spectral and angular dependences are not derived from the material properties of the sample and the incident electron velocity. 
In the simplified model, we consider radiation only from the upper surface, which is realized when the sample is sufficiently thick metal for radiation associated with the lower interface to be negligible (while remaining thin enough for the incident electron to transmit through it). 
The angular distribution in Eq.~\eqref{f_model} is chosen only to represent photon emission predominantly in oblique directions into the upper half-space, and the precise peak angle has no particular physical significance in the following discussion. 
Accordingly, although the quantitative boundaries between the regimes discussed below may depend on the detailed form of $\Gamma(\bm{k})$, the underlying mechanism identified in our analysis is expected to be more general.
We also note that the results in Eqs.~\eqref{purity}, \eqref{Delta_x_x}, \eqref{Delta_qx_kx}, and \eqref{purity_z}--\eqref{dist_qx_kx} are valid for a general form of $\Gamma(\bm{k})$. 
Therefore, these results can be applicable also to the cases that radiation from both the upper and lower surfaces and the resulting interference may become important \cite{Yamamoto_etal1996,Stoger-Pollach_etal2019}.

\section{Electron--photon entanglement}
\label{sec:entanglement}

We now quantitatively evaluate the degree of electron--photon entanglement in the scattered state $\ket{\Psi_{\mathrm{sc}}}$.
Experimentally, $\ket{\Psi_{\mathrm{sc}}}$ can be obtained by post-selecting single-photon emission events in $\ket{\Psi_{\mathrm{fin}}}$.
To characterize the entanglement, we employ two measures:
(i) the purity of the reduced density matrix of the electron,
and (ii) the EPR-type correlation criterion for continuous-variable systems proposed in Ref.~\cite{Mancini_etal2002}.
By using these two measures, we classify the parameter space into three distinct regimes, as shown in Fig.~\ref{fig:entanglement_1}.

In the main illustrative evaluation shown in Fig.~\ref{fig:entanglement_1}, we use the longitudinal coherence length $L_\parallel^\mathrm{el} = 1.3~\mathrm{\mu m}$ for the electron beam, which is typical for a field emission gun \cite{TEM_book2009}.

\subsection{Subsystem purity}
\label{sec:purity}

Since the scattered electron--photon state $\ket{\Psi_{\mathrm{sc}}}$ is a pure state,
we can use the subsystem purity $\mathcal{P}_{\mathrm{sc}}$ of the scattered state to quantify the electron--photon entanglement.
$\mathcal{P}_{\mathrm{sc}}$ is defined as
\begin{align}
  \mathcal{P}_{\mathrm{sc}} =
  \mathrm{Tr}_{\mathrm{el}} \left[
    \bigl( \mathrm{Tr}_{\mathrm{ph}} \ketbra{\Psi_\mathrm{sc}}{\Psi_\mathrm{sc}} \bigr)^2
    \right],
\end{align}
where $\mathrm{Tr}_{\mathrm{el}}$ ($\mathrm{Tr}_{\mathrm{ph}}$) denotes the trace over the electron (photon) system.
This quantity satisfies $0 \le \mathcal{P}_{\mathrm{sc}} \le 1$, and $\mathcal{P}_{\mathrm{sc}} = 1$ holds exactly when $\ket{\Psi_\mathrm{sc}}$ is separable between the electron and photon.
The inverse of the subsystem purity, $K_\mathrm{Schmidt} = 1 / \mathcal{P}_{\mathrm{sc}}$, corresponds to the Schmidt number,
which is widely used to quantify entanglement in continuous-variable bipartite pure states \cite{Grobe_etal1994,Chan_etal2002,Chan_etal2003,LawEberly2004,Fedorov_etal2005,GuoGuo2006,GuoGuo2007,Pires_etal2009,DErricoKarimi2026}.
$K_\mathrm{Schmidt}$ represents the effective number of coefficients in the Schmidt decomposition of $\ket{\Psi_\mathrm{sc}}$.
Thus, a larger $K_\mathrm{Schmidt}$ indicates that a greater number of modes significantly contribute to the entanglement in $\ket{\Psi_\mathrm{sc}}$.
In this sense, a lower $\mathcal{P}_{\mathrm{sc}}$ signals a higher degree of entanglement.

To distinguish strongly entangled regions from a nearly separable region in the $\sigma_{q,\perp}^\mathrm{el}$--$\sigma_{k}^\mathrm{ph}$ parameter space shown in Fig.~\ref{fig:entanglement_1}(a), we adopt $\mathcal{P}_{\mathrm{sc}} = 2/3$ ($K_\mathrm{Schmidt} = 1.5$) as a practical threshold, although $\mathcal{P}_{\mathrm{sc}} = 1$ is the exact condition for separability. This threshold is introduced only for classification and visualization. The corresponding Schmidt number $K_\mathrm{Schmidt} = 1.5$ lies between $K_\mathrm{Schmidt} = 1$ for a separable state and $K_\mathrm{Schmidt} = 2$ for a state with two equally weighted Schmidt modes. The qualitative features of Fig.~\ref{fig:entanglement_1}(a) and our conclusions remain unchanged if the threshold is chosen closer to unity.

Using Eqs.~\eqref{Psi_scattered} and \eqref{psi_sc} for $\ket{\Psi_\mathrm{sc}}$ with Eqs.~\eqref{initial_psi_xy} and \eqref{initial_psi_z} for $\psi_{\mathrm{ini}}$,
we find that $\mathcal{P}_{\mathrm{sc}}$ is given by
\begin{align}
  \mathcal{P}_{\mathrm{sc}}
   & = \int d^3k \, d^3k' \, \Gamma(\bm{k}) \Gamma(\bm{k}')
  \notag
  \\
   & \quad \times \exp[ -\frac{|\bm{k}_\perp - \bm{k}'_\perp|^2}{4 (\sigma_{q,\perp}^\mathrm{el})^2} - \frac{c^2(k - k')^2}{4 v_z^2(\sigma_{q,\parallel}^\mathrm{el})^2} ].
  \label{purity}
\end{align}
We note that this is independent of the undetermined phase $\eta(\bm{k})$.
The derivation of this result is given in Appendix~\ref{appendix:purity}.

We evaluate $\mathcal{P}_{\mathrm{sc}}$ by substituting the model for $\Gamma(\bm{k})$ [Eqs.~\eqref{Gamma_model}--\eqref{f_model}] into the above expression and numerically computing the integral.
In Fig.~\ref{fig:entanglement_1}(b), we show a color plot of $\mathcal{P}_{\mathrm{sc}}$ as a function of $\sigma_{q,\perp}^\mathrm{el}$ and $\sigma_{k}^\mathrm{ph}$ ($L_\perp^\mathrm{el}$ and $\Delta \lambda_\mathrm{ph}$)
for the case of $L_\parallel^\mathrm{el} = 1.3~\mathrm{\mu m}$ ($\sigma_{q,\parallel}^\mathrm{el} \simeq 4.83~\mathrm{\mu m^{-1}}$) and $\lambda_c = 0.5~\mathrm{\mu m}$ ($k_c \simeq 12.6~\mathrm{\mu m^{-1}}$).
We find that
$\mathcal{P}_{\mathrm{sc}} \simeq 1$ in the region satisfying $\sigma_{q,\perp}^\mathrm{el} \gtrsim 20~\mathrm{\mu m^{-1}}$ ($L_\perp^\mathrm{el} \lesssim 0.3~\mathrm{\mu m}$) and $\sigma_{k}^\mathrm{ph} \lesssim 2~\mathrm{\mu m^{-1}}$ ($\Delta \lambda_\mathrm{ph} \lesssim 0.1~\mathrm{\mu m}$),
whereas $\mathcal{P}_{\mathrm{sc}} \simeq 0$ far outside of this region.
The latter indicates that essentially an unbounded number of Schmidt modes contribute to the entanglement,
which is characteristic of entangled states in continuous-variable systems.
This result shows that, for the parameters considered here, strong electron--photon entanglement is realized when $L_\perp^\mathrm{el}$ exceeds several micrometers, a range accessible with highly coherent and collimated electron beams. Increasing $L_\perp^\mathrm{el}$ narrows the transverse momentum distribution of the incident electron and makes the electron--photon momentum correlation clearer, resulting in stronger entanglement.

\subsection{EPR correlation}
\label{sec:EPR}

We next analyze the electron--photon entanglement in the $x$ direction in the scattered state $\ket{\Psi_\mathrm{sc}}$ using the criterion proposed in Ref.~\cite{Mancini_etal2002}, which is employed to detect EPR correlations in continuous-variable systems \cite{Howell_etal2004, DAngelo_etal2005, Walborn_etal2010}.
In the context of the present electron--photon system,
this criterion is based on the $x$-directional uncertainty product of
the relative position, $x_\mathrm{el} - x_\mathrm{ph}$,
and the total wavevector, $q_x + k_x$:
\begin{align}
  \mathcal{D}_{\mathrm{sc}}^2 =
  \expval{[\Delta (x_\mathrm{el} - x_\mathrm{ph})]^2} \expval{[\Delta (q_x + k_x)]^2}.
\end{align}
Here the expectation values are taken in the scattered state $\ket{\Psi_\mathrm{sc}}$.
The criterion states that $\mathcal{D}_{\mathrm{sc}}^2 \ge 1$ for states separable between the electron and photon; thus, $\mathcal{D}_{\mathrm{sc}}^2 < 1$ certifies electron--photon entanglement. 
Furthermore, according to the stricter criterion in Ref.~\cite{Mancini_etal2002}, $\mathcal{D}_{\mathrm{sc}}^2 < 1/4$ certifies an EPR correlation.
Indeed, this type of criterion was recently used in the verification of electron--photon entanglement \cite{Preimesberger_etal2025}.
We therefore adopt $\mathcal{D}_{\mathrm{sc}}^2 = 1/4$ as the threshold for the classification shown in Fig.~\ref{fig:entanglement_1}(a).

Using Eqs.~\eqref{Psi_scattered} and \eqref{psi_sc} for $\ket{\Psi_\mathrm{sc}}$, 
which includes $\psi_\mathrm{ini}(\bm{q}_\perp + \bm{k}_\perp, q_z + ck/v_z)$, 
with Eqs.~\eqref{initial_psi_xy} and \eqref{initial_psi_z} for $\psi_{\mathrm{ini}}$,
we find that the relative position uncertainty is given by
\begin{align}
   & \expval{[\Delta (x_\mathrm{el} - x_\mathrm{ph})]^2}
  \notag
  \\
   & = \frac{1}{8} \int d^3k \left\{ \frac{1}{\Gamma(\bm{k})} \left[ \biggl(\frac{\partial \Gamma(\bm{k})}{\partial k_x} \biggr)^2 + \biggl(\frac{\partial \Gamma(\bm{k})}{\partial k_y} \biggr)^2 \right] \right.
  \notag
  \\
   & \qquad\qquad \left. \quad - 2 \biggl( \frac{\partial^2 \Gamma(\bm{k})}{\partial k_x^2} + \frac{\partial^2 \Gamma(\bm{k})}{\partial k_y^2} \biggr)
  + \frac{c^2 k_\perp^2 \Gamma(\bm{k})}{v_z^2 (\sigma_{q,\parallel}^\mathrm{el})^2 k^2} \right\}
  \notag
  \\
   & \quad + \frac{1}{2} \int d^3k \, \Gamma(\bm{k}) \left[ \Bigl( \frac{\partial \eta(\bm{k})}{\partial k_x} \Bigr)^2 + \Bigl( \frac{\partial \eta(\bm{k})}{\partial k_y} \Bigr)^2 \right],
  \label{Delta_x_x}
\end{align}
and the total wavevector uncertainty is given by
\begin{align}
  \expval{[\Delta (q_x + k_x)]^2} = \bigl( \sigma_{q,\perp}^\mathrm{el} \bigr)^2.
  \label{Delta_qx_kx}
\end{align}
We note that $\expval{[\Delta (q_x + k_x)]^2}$ is independent of both the undetermined phase $\eta(\bm{k})$ and the CL spectrum $\Gamma(\bm{k})$,
whereas $\expval{[\Delta (x_\mathrm{el} - x_\mathrm{ph})]^2}$ depends on them.
The derivation of Eqs.~\eqref{Delta_x_x} and \eqref{Delta_qx_kx} is given in Appendix~\ref{appendix:uncertainty}.
The result in Eq.~\eqref{Delta_qx_kx}, showing $\expval{[\Delta (q_x + k_x)]^2}$ is determined solely by the transverse coherence of the incident electron, can be understood from the translational invariance in the transverse plane. Since the total transverse momentum $\hbar\bm{q}_\perp + \hbar\bm{k}_\perp$ is conserved in this electron--photon system (i.e., the corresponding total transverse momentum operator commutes with $\hat{H} = \hat{H}_0 + \hat{H}_{\mathrm{int}}$), its variance is also conserved and thus coincides with that of the initial state $\ket{\Psi_\mathrm{ini}}$, namely $( \sigma_{q,\perp}^\mathrm{el} )^2$.

Using the model for $\Gamma(\bm{k})$ [Eqs.~\eqref{Gamma_model}--\eqref{f_model}],
we further proceed with the calculation of $\expval{[\Delta (x_\mathrm{el} - x_\mathrm{ph})]^2}$
to obtain (see Appendix~\ref{appendix:Delta_x_x} for the detailed derivation)
\begin{align}
   & \expval{[\Delta (x_\mathrm{el} - x_\mathrm{ph})]^2}
  \notag
  \\
   & = \frac{\sqrt{2 \pi} \mathcal{N}_g}{56}
  \biggl( 19 \sigma_{k}^\mathrm{ph} + \frac{2 k_c^2}{\sigma_{k}^\mathrm{ph}} \biggr) \biggl[ \erf\biggl(\frac{k_c}{\sqrt{2} \sigma_{k}^\mathrm{ph}} \biggr) + 1 \biggr]
  \notag
  \\
   & \quad + \frac{\mathcal{N}_g}{14} k_c \exp \biggl( -\frac{k_c^2}{2 (\sigma_{k}^\mathrm{ph})^2}\biggr)
  + \frac{c^2}{14 v_z^2 (\sigma_{q,\parallel}^\mathrm{el})^2}
  + D_\eta,
  \label{Delta_x_x_model}
\end{align}
where $D_\eta = (1/2) \int d^3k \, \Gamma(\bm{k}) \sum_{\alpha=x,y} [ \partial \eta(\bm{k}) / \partial k_\alpha ]^2$.
To discuss the entanglement intrinsic to the scattering state $\ket{\Psi_\mathrm{sc}}$, it is sufficient to consider the case where the $\eta(\bm{k})$-dependent term $D_\eta$ is minimized,
since the phase $\eta(\bm{k})$ in $\ket{\Psi_\mathrm{sc}}$ can be changed by a local unitary transformation on the photon.
In the following, we focus on the case where $D_\eta = 0$, which yields the minimum.
In practical experiments, however, the effect of the phase $\eta(\bm{k})$ may not be perfectly eliminated; its influence on the uncertainty product $\mathcal{D}_{\mathrm{sc}}^2$ is therefore discussed in Appendix~\ref{appendix:uncertainty_phase}.

We evaluate the uncertainty product $\mathcal{D}_{\mathrm{sc}}^2$ by multiplying Eqs.~\eqref{Delta_qx_kx} and \eqref{Delta_x_x_model}.
In Fig.~\ref{fig:entanglement_1}(c), we show a color plot of $\mathcal{D}_{\mathrm{sc}}^2$ as a function of $\sigma_{q,\perp}^\mathrm{el}$ and $\sigma_{k}^\mathrm{ph}$ ($L_\perp^\mathrm{el}$ and $\Delta \lambda_\mathrm{ph}$)
for the case of $L_\parallel^\mathrm{el} = 1.3~\mathrm{\mu m}$ ($\sigma_{q,\parallel}^\mathrm{el} \simeq 4.83~\mathrm{\mu m^{-1}}$) and $\lambda_c = 0.5~\mathrm{\mu m}$ ($k_c \simeq 12.6~\mathrm{\mu m^{-1}}$).
We find that $\mathcal{D}_{\mathrm{sc}}^2$ is significantly smaller than one in the upper-left region of the plot.
For instance, when $\sigma_{k}^\mathrm{ph} = 3~\mathrm{\mu m^{-1}}$ ($\Delta \lambda_\mathrm{ph} \simeq 0.01~\mathrm{\mu m}$), $L_\perp^\mathrm{el}$ must exceed a few micrometers to achieve strong electron--photon EPR correlation in the $x$ direction.

We also find that, along the $\mathcal{D}_{\mathrm{sc}}^2 = 1/4$ contour shown by the dotted curve in Fig.~\ref{fig:entanglement_1}(c), 
$\sigma_{k}^{\mathrm{ph}} \propto \sigma_{q,\perp}^{\mathrm{el}}$ in the small-$\sigma_{k}^{\mathrm{ph}}$ region [$\sigma_{k}^{\mathrm{ph}} \lesssim 1~\mathrm{\mu m^{-1}}$; the unhatched regime in Fig.~\ref{fig:entanglement_1}(a)], 
whereas $\sigma_{q,\perp}^{\mathrm{el}}$ is approximately constant in the large-$\sigma_{k}^{\mathrm{ph}}$ region [$\sigma_{k}^{\mathrm{ph}} \gtrsim 1~\mathrm{\mu m^{-1}}$; the hatched regime in Fig.~\ref{fig:entanglement_1}(a)].
At the level of the expressions in Eqs.~\eqref{Delta_qx_kx} and \eqref{Delta_x_x_model}, this behavior can be understood as follows.
From Eq.~\eqref{Delta_x_x_model}, when $\sigma_{k}^\mathrm{ph} \ll k_c$ and $\sigma_{k}^\mathrm{ph} \ll \sigma_{q,\parallel}^\mathrm{el}$, the dominant contribution to $\expval{[\Delta (x_\mathrm{el} - x_\mathrm{ph})]^2}$ scales as $1 / (\sigma_{k}^\mathrm{ph})^2$ [the leading term in $\mathcal{N}_g$ for $\sigma_{k}^\mathrm{ph} \ll k_c$ is proportional to $1 / (k_c^2 \sigma_{k}^\mathrm{ph})$; see Eq.~\eqref{N_g}].
Combining this with Eq.~\eqref{Delta_qx_kx}, we obtain $\mathcal{D}_{\mathrm{sc}}^2 \propto (\sigma_{q,\perp}^\mathrm{el} / \sigma_{k}^\mathrm{ph})^2$ in this regime.
This leads to $\sigma_{k}^\mathrm{ph} \propto \sigma_{q,\perp}^\mathrm{el}$ along the line $\mathcal{D}_{\mathrm{sc}}^2 = 1/4$ for small $\sigma_{k}^\mathrm{ph}$.
On the other hand, when $\sigma_{k}^\mathrm{ph} \gg \sigma_{q,\parallel}^\mathrm{el}$, the dominant contribution to $\expval{[\Delta (x_\mathrm{el} - x_\mathrm{ph})]^2}$ scales as $1 / (\sigma_{q,\parallel}^\mathrm{el})^2$.
Combining this with Eq.~\eqref{Delta_qx_kx}, we obtain $\mathcal{D}_{\mathrm{sc}}^2 \propto (\sigma_{q,\perp}^\mathrm{el} / \sigma_{q,\parallel}^\mathrm{el})^2$ in this regime.
This implies $\sigma_{q,\perp}^\mathrm{el} \propto \sigma_{q,\parallel}^\mathrm{el}$ (independent of $\sigma_{k}^\mathrm{ph}$) along the line $\mathcal{D}_{\mathrm{sc}}^2 = 1/4$ for large $\sigma_{k}^\mathrm{ph}$.

\subsection{Wave-like, particle-like, and classical regimes}
\label{sec:regimes}

\begin{figure*}[tb]
  \includegraphics[width=\linewidth]{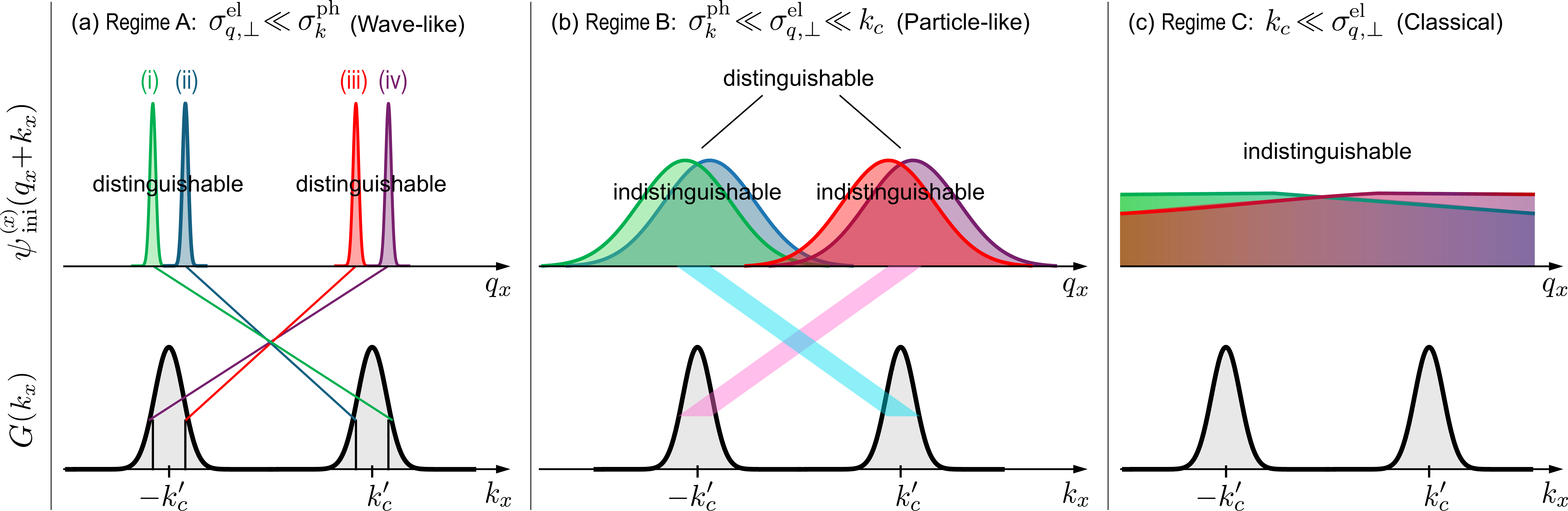}
  \caption{\label{fig:regimes}
  Schematic illustrations of the $x$-directional wavevector-space functions in Eq.~\eqref{rough_state_sc} shown for the three regimes labeled A, B, and C in Fig.~\ref{fig:entanglement_1}(a).
  The upper panels depict $\psi^{(x)}_{\mathrm{ini}}(q_x + k_x)$ for
  (i) $k_x = k_c' + \sigma_{k}^\mathrm{ph}$,
  (ii) $k_x = k_c' - \sigma_{k}^\mathrm{ph}$,
  (iii) $k_x = -k_c' + \sigma_{k}^\mathrm{ph}$, and
  (iv) $k_x = -k_c' - \sigma_{k}^\mathrm{ph}$.
  The lower panels depict $G(k_x)$.
  (a) Regime~A, where $\sigma_{q,\perp}^\mathrm{el} \ll \sigma_{k}^\mathrm{ph}$.
  The functions $\psi^{(x)}_\mathrm{ini}(q_x + k_x)$ of $q_x$ for different $k_x$ are distinguishable within each of the two peaks of $G(k_x)$,
  which leads to nearly one-to-one correlation (represented by thin lines): for each photon with $k_x$, a corresponding electron with $q_x = -k_x$ is correlated.
  (b) Regime~B, where $\sigma_{k}^\mathrm{ph} \ll \sigma_{q,\perp}^\mathrm{el} \ll k_c$.
  The functions $\psi^{(x)}_\mathrm{ini}(q_x + k_x)$ for different $k_x$ are indistinguishable within each peak of $G(k_x)$ but distinguishable between the two peaks,
  which leads to less specific correlations (represented by thick lines): photons traveling in the positive direction tend to correlate with electrons traveling in the negative direction, and vice versa, but without a sharp one-to-one correspondence.
  (c) Regime~C, where $k_c \ll \sigma_{q,\perp}^\mathrm{el}$.
  The functions $\psi^{(x)}_\mathrm{ini}(q_x + k_x)$ for different $k_x$ are indistinguishable even between the two peaks of $G(k_x)$.
  As explained in the main text, Regimes A, B, and C corresponds to the wave-like, particle-like, and classical regimes, respectively.
  }
\end{figure*}

By comparing the results for $\mathcal{P}_{\mathrm{sc}}$ in Fig.~\ref{fig:entanglement_1}(b) and $\mathcal{D}_{\mathrm{sc}}^2$ in Fig.~\ref{fig:entanglement_1}(c), we identify three distinct parameter regimes, labeled A, B, and C in Fig.~\ref{fig:entanglement_1}(a).
Regime~A corresponds to strong entanglement according to both measures.
Regime~B exhibits strong entanglement in terms of $\mathcal{P}_{\mathrm{sc}}$ but not according to $\mathcal{D}_{\mathrm{sc}}^2$.
Regime~C corresponds to almost separable (non-entangled) regime.
As we will discuss in Sec.~\ref{sec:longitudinal}, this parameter space is also divided into small- and large-$\sigma_{k}^\mathrm{ph}$ regimes, which are distinguished by the hatching in Fig.~\ref{fig:entanglement_1}(a).
In this subsection, we focus on the unhatched (small-$\sigma_{k}^\mathrm{ph}$) regime in Fig.~\ref{fig:entanglement_1}(a) and provide a qualitative understanding of Regimes A, B, and C.
We show that these regimes can be identified with the wave-like, particle-like, and classical regimes.
While the earlier work \cite{Huang_etal2023} formulated this classification in frequency (energy) space, we extend it here to wavevector (momentum) space and present the typical forms of the electron--photon state in each regime.

For the qualitative understanding, it is useful to focus on the entanglement between the $x$-directional degrees of freedom of the electron and the photon. 
Starting from the scattered state $\ket{\Psi_{\mathrm{sc}}}$ in Eqs.~\eqref{Psi_scattered}, \eqref{psi_sc},
we introduced, at a qualitative level, the $x$-directional state
\begin{align}
  \ket{\Psi_{\mathrm{sc},x}} \approx
  \int dq_x \, dk_x \, \psi^{(x)}_\mathrm{ini}(q_x + k_x) \, G(k_x) \ket{q_x}_\mathrm{el} \ket{k_x}_\mathrm{ph}.
  \label{rough_state_sc}
\end{align}
Here, $G(k_x)$ represents an effective projection of $\sqrt{\Gamma(\bm{k})}$ onto the $k_x$ direction, and $\ket{q_x}_\mathrm{el}$ ($\ket{k_x}_\mathrm{ph}$) denotes the electron's (photon's) wavevector state in the $x$ direction. 
Since $\sqrt{\Gamma(\bm{k})}$ is peaked around $|\bm{k}| = k_c$ with the width $\sigma_{k}^\mathrm{ph}$, the effective function $G(k_x)$ has two peaks at $k_x = \pm k_c'$ with almost the same widths, where $k_c'$ has the same order of magnitude as $k_c$.
When $\sigma_{k}^\mathrm{ph} \ll k_c$, the two peaks are well separated, as illustrated in the lower panels of Fig.~\ref{fig:regimes}.
Strictly speaking, $\ket{\Psi_{\mathrm{sc}}}$ cannot be reduced to $\ket{\Psi_{\mathrm{sc},x}}$, since the electron's $x$-directional degree of freedom is entangled not only with the photonic $x$-directional degree of freedom but also with all photonic degrees of freedom and the electron's $y$-directional degree of freedom.
Nevertheless, Eq.~\eqref{rough_state_sc} is sufficient for the following qualitative discussion.
From Eq.~\eqref{rough_state_sc}, we see that a photon with $x$-directional wavenumber $k_x$ is correlated with an electron wave packet described by $\psi^{(x)}_\mathrm{ini}(q_x + k_x)$, which is centered around $q_x=-k_x$ with the width $\sigma_{q,\perp}^\mathrm{el}$.
Such correlations are depicted as the crossing lines between the upper and lower panels in Fig.~\ref{fig:regimes}.

In Regime~A, since $\sigma_{q,\perp}^\mathrm{el} \ll \sigma_{k}^\mathrm{ph}$,
the Gaussian function $\psi^{(x)}_\mathrm{ini}(q_x + k_x)$ can be approximated as a delta-like function within each of the two peaks of $G(k_x)$ in wavevector space, as schematically illustrated in the upper panel of Fig.~\ref{fig:regimes}(a).
This sharp behavior in the wave-like regime was also discussed in frequency space \cite{Huang_etal2023}.
As a result, Eq.~\eqref{rough_state_sc} reduces to an EPR-like state,
\begin{align*}
  \ket{\Psi_{\mathrm{sc},x}} \approx \int dk_x \, G(k_x) \ket{-k_x}_\mathrm{el} \ket{k_x}_\mathrm{ph}.
\end{align*}
Therefore, the entanglement is detected both by $\mathcal{P}_{\mathrm{sc}}$ and by $\mathcal{D}_{\mathrm{sc}}^2$ in this regime.
As seen in this equation, a photon with $k_x$ is correlated with an electron with $q_x \approx -k_x$, which is schematically depicted as thin crossing lines in Fig.~\ref{fig:regimes}(a).
Moreover, if the photonic state is known to be $\ket{k_x}_\mathrm{ph}$, the electron state is approximately a plane-wave state, $\ket{-k_x}_\mathrm{el}$.
In this sense, Regime~A corresponds to the wave-like regime.

In Regime~B, since $\sigma_{k}^\mathrm{ph} \ll \sigma_{q,\perp}^\mathrm{el} \ll k_c$,
the variation of $\psi^{(x)}_\mathrm{ini}(q_x + k_x)$ for different $k_x$ cannot be resolved within each peak of $G(k_x)$ in wavevector space, as schematically illustrated in Fig.~\ref{fig:regimes}(b).
This broad behavior in the particle-like regime was also discussed in frequency space \cite{Huang_etal2023}.
Consequently, Eq.~\eqref{rough_state_sc} can be further approximated as
\begin{align*}
  \ket{\Psi_{\mathrm{sc},x}}
   & \approx \int dq_x \, \psi^{(x)}_\mathrm{ini}(q_x + k_c') \ket{q_x}_\mathrm{el} \int_0^\infty dk_x \, G(k_x) \ket{k_x}_\mathrm{ph}
  \\
   & + \int dq_x \, \psi^{(x)}_\mathrm{ini}(q_x - k_c') \ket{q_x}_\mathrm{el} \int_{-\infty}^0 dk_x \, G(k_x) \ket{k_x}_\mathrm{ph}.
\end{align*}
This represents a superposition of two distinct states:
a photon with $k_x$ around $-k_c' \pm \sigma_{k}^\mathrm{ph}$ is correlated with an electron in the state $\psi^{(x)}_\mathrm{ini}(q_x - k_c')$, 
and a photon with $k_x$ around $k_c' \pm \sigma_{k}^\mathrm{ph}$ is correlated with an electron in the state $\psi^{(x)}_\mathrm{ini}(q_x + k_c')$.
Such correlations are schematically depicted as thick crossing lines in Fig.~\ref{fig:regimes}(b).
We note that in the full two-dimensional transverse space, the electron and photon are also entangled in the direction of azimuthal angle $\phi$.
Since the angle $\phi$ is a continuous variable, there remain nearly infinitely many Schmidt modes, and therefore $\mathcal{P}_{\mathrm{sc}} \simeq 0$, even though $\mathcal{D}_{\mathrm{sc}}^2$ does not detect entanglement in this regime.
Moreover, if the photonic state is known to be $\ket{k_x}_\mathrm{ph}$, the electron state becomes a wave-packet (particle-like) state approximately given by $\int dq_x \, \psi^{(x)}_\mathrm{ini} \bigl( q_x + \mathrm{sgn}(k_x) k_c \bigr) \ket{q_x}_\mathrm{el}$.
In this sense, Regime~B corresponds to the particle-like regime.

In Regime~C, where $k_c \ll \sigma_{q,\perp}^\mathrm{el}$,
the function $\psi^{(x)}_\mathrm{ini}(q_x + k_x)$ for different $k_x$ cannot be resolved even between $k_x = k_c$ and $k_x = -k_c$, as schematically illustrated in Fig.~\ref{fig:regimes}(c).
Accordingly, Eq.~\eqref{rough_state_sc} reduces to
\begin{align*}
  \ket{\Psi_{\mathrm{sc},x}} \approx
  \int dq_x \, \psi^{(x)}_\mathrm{ini}(q_x) \ket{q_x}_\mathrm{el} \int dk_x \, G(k_x) \ket{k_x}_\mathrm{ph}.
\end{align*}
This is a separable state, leading to $\mathcal{P}_{\mathrm{sc}} \simeq 1$ and $\mathcal{D}_{\mathrm{sc}}^2 \ge 1$.
Therefore, this regime corresponds to the classical regime.

In summary, in the small-$\sigma_{k}^\mathrm{ph}$ regime ($\sigma_{k}^\mathrm{ph} \ll \sigma_{q,\parallel}^\mathrm{el}$ and $\sigma_{k}^\mathrm{ph} \ll k_c$),
the boundary between Regimes A and B lies around $\sigma_{q,\perp}^\mathrm{el} \approx \sigma_{k}^\mathrm{ph}$, while that between Regimes B and C lies around $\sigma_{q,\perp}^\mathrm{el} \approx k_c$.

\begin{figure}[tb]
  \includegraphics[width=\linewidth]{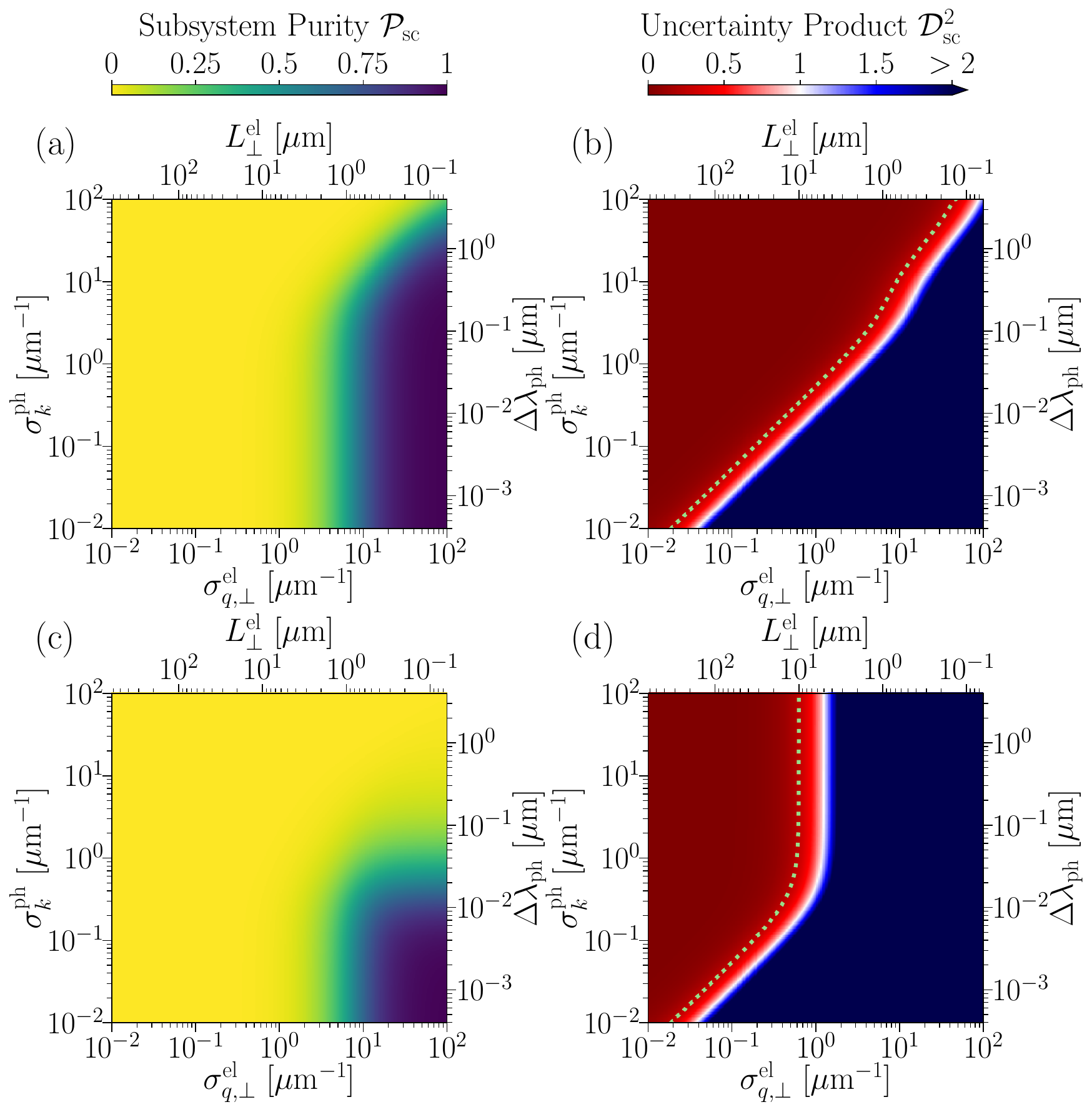}
  \caption{\label{fig:entanglement_2}
  Entanglement measures of the scattered state $\ket{\Psi_{\mathrm{sc}}}$ for different values of $L_\parallel^\mathrm{el}$, plotted as functions of $\sigma_{q,\perp}^\mathrm{el}$ and $\sigma_{k}^\mathrm{ph}$.
  Left panels [(a) and (c)] show the results for the subsystem purity $\mathcal{P}_{\mathrm{sc}}$, and right panels [(b) and (d)] show those for the uncertainty product $\mathcal{D}_{\mathrm{sc}}^2$.
  (a) and (b): Results for $L_\parallel^\mathrm{el} = 0.13~\mathrm{\mu m}$ ($\sigma_{q,\parallel}^\mathrm{el} \simeq 48.3~\mathrm{\mu m^{-1}}$).
  (c) and (d): Results for $L_\parallel^\mathrm{el} = 13~\mathrm{\mu m}$ ($\sigma_{q,\parallel}^\mathrm{el} \simeq 0.483~\mathrm{\mu m^{-1}}$).
  The other parameter values are the same as those in Fig.~\ref{fig:entanglement_1}.
  The dotted curves in (b) and (d) represent the EPR-correlation threshold $\mathcal{D}_{\mathrm{sc}}^2 = 1/4$.
  }
\end{figure}

\subsection{Influence of electron longitudinal coherence}
\label{sec:longitudinal}

We next examine how the longitudinal coherence of the electron beam affects the entanglement.
To clearly illustrate the dependence on the longitudinal coherence length $L_\parallel^{\rm el}$, 
we consider values of $L_\parallel^{\rm el}$ that are one order of magnitude smaller and larger than the value used in Fig.~\ref{fig:entanglement_1} and the preceding subsections.

Figures~\ref{fig:entanglement_2}(a) and (c) show the subsystem purity $\mathcal{P}_{\mathrm{sc}}$ for different values of $L_\parallel^\mathrm{el}$.
Compared with Fig.~\ref{fig:entanglement_1}(b),
as $L_\parallel^\mathrm{el}$ increases ($\sigma_{q,\parallel}^\mathrm{el}$ decreases),
the region where $\mathcal{P}_{\mathrm{sc}}$ indicates strong entanglement expands toward Regime C [upper right of Fig.~\ref{fig:entanglement_2}(c)].
Figures~\ref{fig:entanglement_2}(b) and (d) show the corresponding uncertainty product $\mathcal{D}_{\mathrm{sc}}^{2}$.
Compared with Fig.~\ref{fig:entanglement_1}(c),
as $L_\parallel^\mathrm{el}$ increases ($\sigma_{q,\parallel}^\mathrm{el}$ decreases),
the region where $\mathcal{D}_{\mathrm{sc}}^{2}$ fails to detect entanglement grows toward Regime A.
This behavior can be understood, at the level of Eq.~\eqref{Delta_x_x_model}, as arising from the term $c^2 / [14 v_z^2 (\sigma_{q,\parallel}^\mathrm{el})^2]$.

The origin of these seemingly opposite trends of $\mathcal{P}_{\mathrm{sc}}$ and $\mathcal{D}_{\mathrm{sc}}^{2}$ can be explained as follows.
As $L_\parallel^\mathrm{el}$ increases,
the width $\sigma_{q,\parallel}^\mathrm{el}$ of $\psi^{(z)}_\mathrm{ini}(q_z + ck/v_z)$ in $\ket{\Psi_{\mathrm{sc}}}$ [Eq.~\eqref{Psi_scattered} with Eq.~\eqref{psi_sc}] becomes narrower.
When $\sigma_{q,\parallel}^\mathrm{el} < \sigma_{k}^\mathrm{ph}$, the entanglement between the electron's longitudinal ($z$-directional) degree of freedom and the photonic degrees of freedom becomes significant in a manner similar to that discussed in Sec.~\ref{sec:regimes}.
Consequently, the electron--photon entanglement detected by $\mathcal{P}_{\mathrm{sc}}$ is present in the region $\sigma_{k}^\mathrm{ph} > \sigma_{q,\parallel}^\mathrm{el}$, and the region extends as $L_\parallel^\mathrm{el}$ increases [Figs.~\ref{fig:entanglement_1}(b) and \ref{fig:entanglement_2}(c)].

Since the argument of $\psi^{(z)}_\mathrm{ini}(q_z + ck/v_z)$ depends on $q_z$ and $k_\perp$,
the electron's longitudinal degree of freedom becomes entangled with the photonic transverse degrees of freedom for $\sigma_{k}^\mathrm{ph} > \sigma_{q,\parallel}^\mathrm{el}$.
Moreover, when $\sigma_{q,\parallel}^\mathrm{el} < \sigma_{q,\perp}^\mathrm{el}$ in addition to $\sigma_{q,\parallel}^\mathrm{el} < \sigma_{k}^\mathrm{ph}$, the width of $\psi^{(z)}_\mathrm{ini}(q_z + ck/v_z)$ becomes smaller than that of $\psi^{(x)}_\mathrm{ini}(q_x + k_x)$.
As a result, the quantum correlation involving the electron's longitudinal degree of freedom becomes stronger than that involving the electron's transverse degrees of freedom.
Consequently, the transverse EPR correlation between the electron's and photon's $x$-degrees of freedom leaks into correlations involving the electron's $z$-degree of freedom. This weakens the transverse EPR correlation, causing $\mathcal{D}_{\mathrm{sc}}^{2}$ to fail to detect entanglement.
Indeed, in the upper region ($\sigma_{k}^\mathrm{ph} > \sigma_{q,\parallel}^\mathrm{el}$) in Figs.~\ref{fig:entanglement_1}(c) and \ref{fig:entanglement_2}(d), $\mathcal{D}_{\mathrm{sc}}^{2} > 1$ for $\sigma_{q,\perp}^\mathrm{el} \gtrsim 2 \sigma_{q,\parallel}^\mathrm{el}$.

\begin{figure}[tb]
  \includegraphics[width=.8\linewidth]{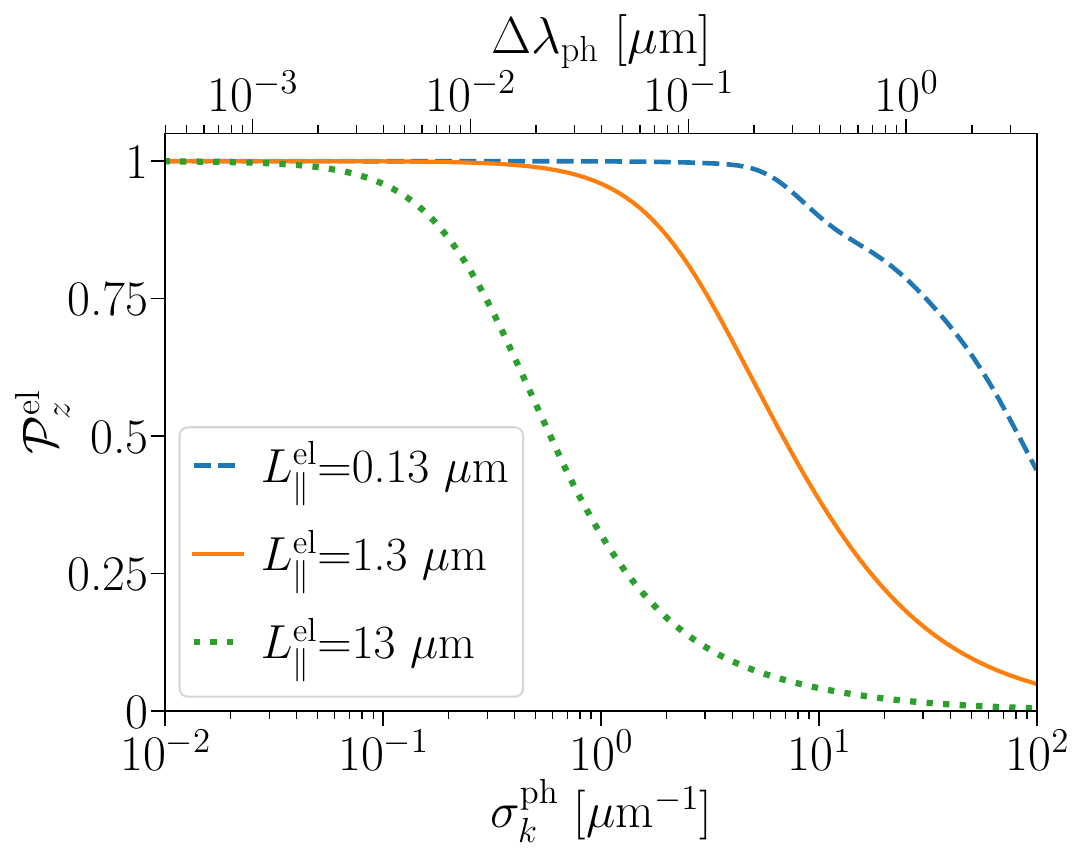}
  \caption{\label{fig:purity_z}
  Subsystem purity $\mathcal{P}^{\mathrm{el}}_{z}$ of the electron's $z$-degree of freedom in the scattered state $\ket{\Psi_{\mathrm{sc}}}$ as a function of the luminescence spectral width $\sigma_{k}^\mathrm{ph}$.
  The dashed, solid, and dotted curves correspond to $L_\parallel^\mathrm{el} = 0.13, 1.3$ and $13~\mathrm{\mu m}$ ($\sigma_{q,\parallel}^\mathrm{el} \simeq 48.3, 4.83$ and $0.483~\mathrm{\mu m^{-1}}$), respectively.
  The other parameter values are the same as those in Fig.~\ref{fig:entanglement_1}.
  }
\end{figure}

To confirm the behavior of the entanglement involving the electron's longitudinal degree of freedom in the above discussion, we examine the subsystem purity $\mathcal{P}^{\mathrm{el}}_{z}$ of the electron's $z$-degree of freedom.
$\mathcal{P}^{\mathrm{el}}_{z}$ is defined as
\begin{align}
  \mathcal{P}^{\mathrm{el}}_{z} =
  \mathrm{Tr}^{\mathrm{el}}_{z} \left[
    \bigl( \mathrm{Tr}^{\mathrm{el}}_{xy} \mathrm{Tr}_{\mathrm{ph}} \ketbra{\Psi_\mathrm{sc}}{\Psi_\mathrm{sc}} \bigr)^2
    \right],
\end{align}
where $\mathrm{Tr}^{\mathrm{el}}_{z}$ and $\mathrm{Tr}^{\mathrm{el}}_{xy}$ denote the traces over the electron's $z$ and $xy$ degrees of freedom, respectively.
Similarly to $\mathcal{P}_{\mathrm{sc}}$, $\mathcal{P}^{\mathrm{el}}_{z}$ satisfies $0 \le \mathcal{P}^{\mathrm{el}}_{z} \le 1$, and a small value of $\mathcal{P}^{\mathrm{el}}_{z}$ indicates significant entanglement involving the electron's longitudinal degree of freedom.
Using Eqs.~\eqref{Psi_scattered} and \eqref{psi_sc} for $\ket{\Psi_\mathrm{sc}}$ with Eqs.~\eqref{initial_psi_xy} and \eqref{initial_psi_z} for $\psi_{\mathrm{ini}}$
in a manner similar to that in Eq.~\eqref{purity}, we obtain
\begin{align}
  \mathcal{P}^{\mathrm{el}}_{z}
  = \int d^3k \, d^3k' \, \Gamma(\bm{k}) \Gamma(\bm{k}')
  \exp[- \frac{c^2(k - k')^2}{4 v_z^2(\sigma_{q,\parallel}^\mathrm{el})^2} ].
  \label{purity_z}
\end{align}
We note that this is independent of both $\sigma_{q,\perp}^\mathrm{el}$ and $\eta(\bm{k})$.

Figure~\ref{fig:purity_z} shows the numerical results of $\mathcal{P}^{\mathrm{el}}_{z}$ for several values of $L_\parallel^\mathrm{el}$, plotted against $\sigma_{k}^\mathrm{ph}$.
We find that $\mathcal{P}^{\mathrm{el}}_{z}$ starts to decrease from unity around $\sigma_{k}^\mathrm{ph} \approx \sigma_{q,\parallel}^\mathrm{el}$.
This confirms that the electron's longitudinal degree of freedom becomes significantly entangled with the remaining ones for $\sigma_{k}^\mathrm{ph} > \sigma_{q,\parallel}^\mathrm{el}$.
In Fig.~\ref{fig:entanglement_1}(a), to explicitly indicate the region where this type of entanglement is present, we apply hatching to the area with $\mathcal{P}^{\mathrm{el}}_{z} < 2/3$.

Finally in this subsection, we discuss the upper region ($\sigma_{k}^\mathrm{ph} \gtrsim 10~\mathrm{\mu m^{-1}}$) in Figs.~\ref{fig:entanglement_2}(a) and (b), where we observe that $\sigma_{q,\perp}^\mathrm{el} \propto \sigma_{k}^\mathrm{ph}$ along the boundaries in both plots.
In these plots, since $\sigma_{k}^\mathrm{ph} < \sigma_{q,\parallel}^\mathrm{el}$, the electron's longitudinal degree of freedom is approximately separable from the remaining ones.
Therefore, Eq.~\eqref{rough_state_sc} is applicable.
However, since $\sigma_{k}^\mathrm{ph} > k_c$ also holds in the upper region, the two peaks of $G(k_x)$ in Eq.~\eqref{rough_state_sc} merge into a single peak with a width $\sigma_{k}^\mathrm{ph}$.
Consequently, the particle-like regime [corresponding to Fig.~\ref{fig:regimes}(b)] is absent in this situation, and the boundary between the entangled and classical regimes satisfies $\sigma_{q,\perp}^\mathrm{el} \approx \sigma_{k}^\mathrm{ph}$.

\subsection{Electron--photon joint probability distribution}

\begin{figure}[tb]
  \includegraphics[width=\linewidth]{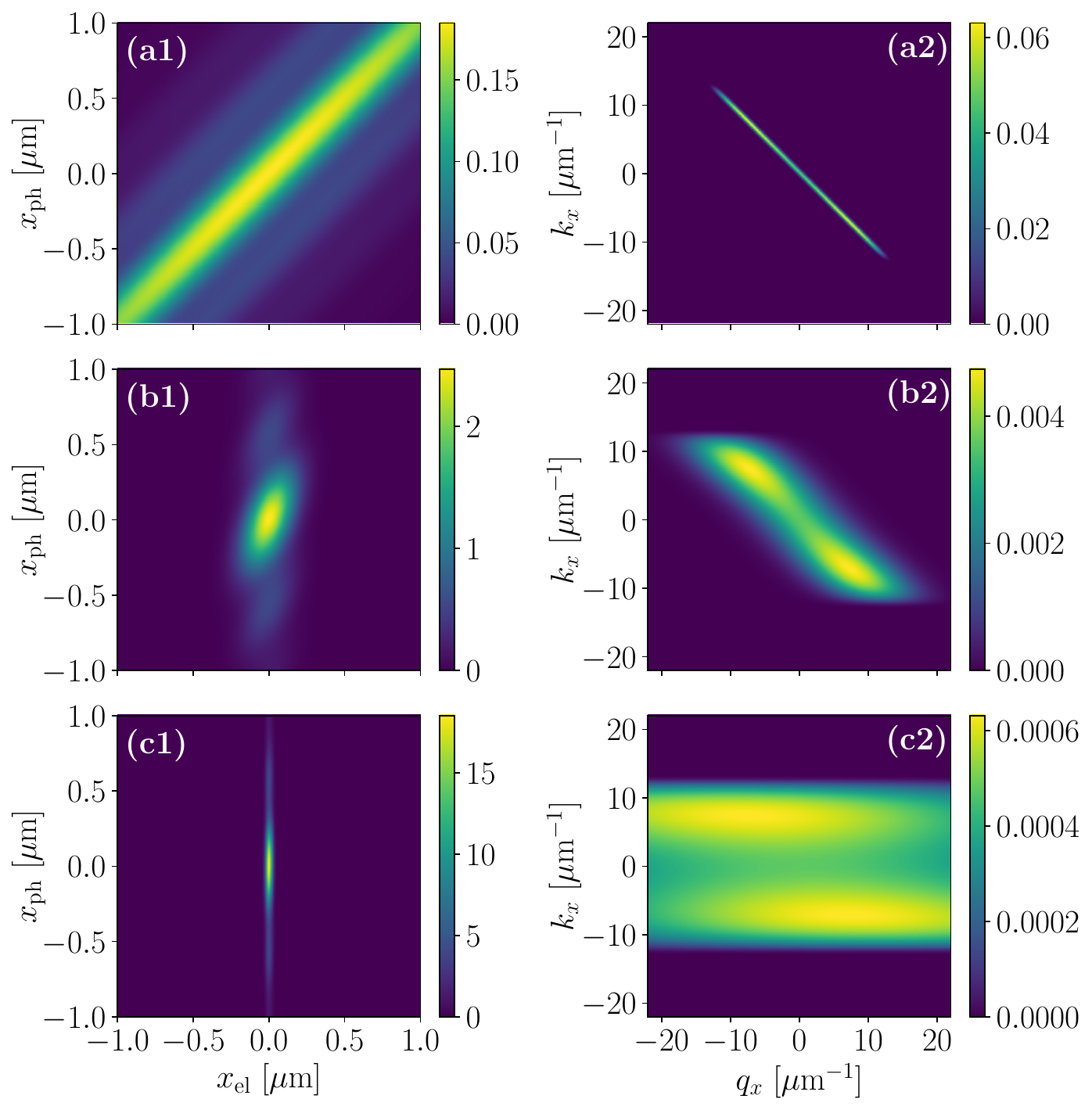}
  \caption{\label{fig:distributions}
    The electron--photon joint probability distributions along the $x$ direction for the states indicated by the cross symbols for $\sigma_{k}^\mathrm{ph} = 0.3~\mathrm{\mu m^{-1}}$ ($\Delta \lambda_{\mathrm{ph}} \simeq 0.119~\mathrm{\mu m}$) in Fig.~\ref{fig:entanglement_1}(a).
    The left panels show the spatial distributions $P(x_\mathrm{el},x_\mathrm{ph})$, and the right panels show the wavevector distributions $P(q_x, k_x)$. (a1) and (a2) correspond to $L_\perp^{\mathrm{el}} = 20~\mathrm{\mu m}$ ($\sigma_{q,\perp}^\mathrm{el} \simeq ~0.314\mathrm{\mu m^{-1}}$), (b1) and (b2) to $L_\perp^{\mathrm{el}} = 1.5~\mathrm{\mu m}$ ($\sigma_{q,\perp}^\mathrm{el} \simeq 4.19~\mathrm{\mu m^{-1}}$), and (c1) and (c2) to $L_\perp^{\mathrm{el}} = 0.2~\mathrm{\mu m}$ ($\sigma_{q,\perp}^\mathrm{el} \simeq 31.4~\mathrm{\mu m^{-1}}$).
    The other parameter values are the same as those in Fig.~\ref{fig:entanglement_1}.
  }
\end{figure}

Finally, we examine the electron--photon joint probability distributions in the scattered state $\ket{\Psi_{\mathrm{sc}}}$ in the three regimes. The distributions can be measured by coincidence counts in experiments.
Using Eqs.~\eqref{Psi_scattered} and \eqref{psi_sc} for $\ket{\Psi_{\mathrm{sc}}}$ and with Eqs.~\eqref{initial_psi_xy} and \eqref{initial_psi_z} for $\psi_{\mathrm{ini}}$,
we find that the joint spatial distribution in the $x$ direction is given by
\begin{align}
  P(x_\mathrm{el},x_\mathrm{ph})
   & = \frac{1}{\sqrt{2\pi^3}}
  \sigma_{q,\perp}^\mathrm{el} e^{-2 (\sigma_{q,\perp}^\mathrm{el})^2 x_\mathrm{el}^2}
  \notag
  \\
   & \times \int dk_x dk_y dk_z dk_x' \sqrt{\Gamma(\bm{k}) \Gamma(\bm{k}')}
  e^{i[\eta(\bm{k}) - \eta(\bm{k}')]}
  \notag
  \\
   & \quad \times
  e^{ -i(k_x - k_x')(x_\mathrm{el} - x_\mathrm{ph}) - c^2 (k - k')^2 / [8 v_z^2 (\sigma_{q,\parallel}^\mathrm{el})^2] },
  \label{dist_x_x}
\end{align}
where $\bm{k} = (k_x, k_y, k_z)$, $\bm{k}' = (k_x', k_y, k_z)$, $k = |\bm{k}|$, and $k' = |\bm{k}'|$
(see Appendix~\ref{appendix:dist_x_x} for the derivation of this equation).
We also find that the joint wavevector (momentum) distribution in the $x$ direction is given by
\begin{align}
  P(q_x, k_x)
  = \big| \psi^{(x)}_\mathrm{ini}(q_x + k_x) \big|^2 
  \int dk_y \, dk_z \, \Gamma(\bm{k}),
  \label{dist_qx_kx}
\end{align}
where $q_x$ is the $x$ component of the scattered-electron wavevector and $\bm{k} = (k_x, k_y, k_z)$ is the emitted photon wavevector
(see Appendix~\ref{appendix:dist_qx_kx} for the derivation of this equation).
We note that $P(q_x, k_x)$ is independent of the undetermined phase $\eta(\bm{k})$.

We evaluate these distributions by substituting the model for $\Gamma(\bm{k})$ [Eqs.~\eqref{Gamma_model}--\eqref{f_model}] into the above expressions and numerically computing the integrals.
For the evaluation of $P(x_\mathrm{el},x_\mathrm{ph})$, we neglect the phase factor $e^{i[\eta(\bm{k}) - \eta(\bm{k}')]}$ in Eq.~\eqref{dist_x_x}, assuming that the phase in $\ket{\Psi_{\mathrm{sc}}}$ is eliminated by a local operation on the photon.

Figure~\ref{fig:distributions} shows the results: the left panels display the spatial distributions and the right panels display the wavevector distributions.
In Regime~A (wave-like regime), both the spatial and wavevector distributions exhibit strong correlations, as seen in Figs.~\ref{fig:distributions}(a1) and \ref{fig:distributions}(a2).
In Regime~B (particle-like regime), the spatial correlations are largely washed out, as seen in Fig.~\ref{fig:distributions}(b1), whereas the wavevector distribution retains a characteristic bimodal structure corresponding to $k_x \simeq \pm k_c$, as shown in Fig.~\ref{fig:distributions}(b2).
This structure reflects the two-component entangled state discussed in Sec.~\ref{sec:regimes}.
In Regime~C (classical regime), both the spatial and wavevector joint distributions exhibit no noticeable correlations, as shown in Figs.~\ref{fig:distributions}(c1) and \ref{fig:distributions}(c2).
This confirms that the scattered state has almost no entanglement in this regime.

\section{Conclusion}
\label{sec:conclusion}

We have developed a theoretical framework to describe the quantum state of electron--photon pairs generated in coherent cathodoluminescence (CL), in which the scattered state $\ket{\Psi_{\mathrm{sc}}}$ is reconstructed from the experimentally accessible luminescence spectrum.
Within this framework, we have quantified the entanglement of the scattered state using the subsystem purities $\mathcal{P}_{\mathrm{sc}}$ and $\mathcal{P}_z^{\mathrm{el}}$, together with the EPR-type uncertainty product $\mathcal{D}^2_{\mathrm{sc}}$.
These measures reveal three distinct regimes---wave-like, particle-like, and classical---each characterized by qualitatively different structures of transverse quantum correlations.
We have shown that these regimes are governed by the interplay between the transverse and longitudinal coherence of the electron beam and the spectral width of the emitted photons.
Our analysis further demonstrates that increasing the longitudinal coherence of the electron beam induces additional correlations involving the longitudinal degree of freedom, which in turn degrade the transverse EPR correlations.
This result indicates that a relatively short longitudinal coherence is favorable for generating strongly entangled electron--photon pairs with pronounced transverse quantum correlations.

The framework assumes translational invariance in the plane perpendicular to the electron beam. This condition can be realized, for example, for transition radiation at a planar interface. Aloof illumination represents another potentially relevant geometry, but it does not generally satisfy the transverse translational symmetry assumed in the present formulation. Describing such a setting would therefore require deriving the electron--photon scattering state from the electromagnetic response specific to the sample geometry. Investigating electron--photon entanglement in aloof, non-planar, or laterally structured geometries, including Smith--Purcell radiation and certain configurations of Cherenkov radiation, remains a subject for future work.

Our analysis also assumes that, conditioned on photon emission, the scattered electron--photon state is pure. In realistic samples, competing inelastic channels may reduce the probability of selecting the desired radiative electron--photon events. Moreover, unobserved nonradiative or environmental degrees of freedom may render the conditional state mixed and reduce the observable entanglement. Incorporating such effects is also left for future work.

Our findings provide a unified perspective on quantum correlations in CL and offer practical guidelines for engineering entangled electron--photon states. 
In particular, the electron--photon momentum correlations considered here can be probed through angle-resolved coincidence detection of the scattered electron and emitted photon, while the conditions identified in this work, including a sufficiently large transverse coherence length of the incident electron, provide guidance for realizing strong entanglement. These results open new possibilities for quantum imaging, quantum sensing, conditional state preparation, and other emerging quantum-enabled techniques in electron microscopy.

\begin{acknowledgments}
  This work was supported by JST CREST Grant No. JPMJCR25I3 and JSPS KAKENHI Grants No. JP18K03454, No. JP21K18195, No. JP22H01963, No. JP23K23231, and No. JP22H05032. One of the authors (K.A.) is grateful to Dr. Takeshi Ohshima for his valuable support for the research.
\end{acknowledgments}

\appendix

\section{Derivation of the scattered state}
\label{appendix:derivation_psi_sc}

In this appendix, we derive the scattered state $\ket{\Psi_{\mathrm{sc}}}$ in Eq.~\eqref{Psi_scattered} with Eq.~\eqref{psi_sc}.
In the derivation, we use a perturbation theory in the electron--photon interaction $\hat{H}_\mathrm{int}$ and the in-plane momentum conservation condition [Eq.~\eqref{in-plane-momentum-conservation}] for $\hat{H}_\mathrm{int}$.

\subsection{Perturbation theory}
\label{appendix:perturbation}

In this paper, we consider the case in which each electron in the electron beam induces at most one photon emission, if it occurs.
In general, a state of the electron--photon system whose photon number is zero or one is expressed as
\begin{align}
  \ket{\Psi(t)}
   & = \int d^3q \, \alpha_0(t, \bm{q}) e^{-i\varepsilon_{\bm{q}} t} \ket{\bm{q}, \mathrm{vac}}
  \notag
  \\
   & \quad + \int d^3q \, d^3k \, \alpha_1(t, \bm{q}, \bm{k}) e^{-i(\varepsilon_{\bm{q}} + \omega_{\bm{k}}) t } \ket{\bm{q}, \bm{k}},
  \label{Psi_t}
\end{align}
where $\varepsilon_{\bm{q}} = \hbar q^2 / (2m)$ and $\omega_{\bm{k}} = ck$.
Since the initial state is $\ket{\Psi_\mathrm{ini}} = \int d^3 q_{\mathrm{in}} \, \psi_\mathrm{ini}(\bm{q}_{\mathrm{in}}) \ket{\bm{q}_{\mathrm{in}}, \mathrm{vac}}$,
the initial condition for the amplitudes $\alpha_0(t, \bm{q})$ and $\alpha_1(t, \bm{q}, \bm{k})$ are given by
\begin{align}
   & \alpha_{0,\mathrm{ini}}(\bm{q}) = \psi_\mathrm{ini}(\bm{q}),
  \label{initial_alpha0}
  \\
   & \alpha_{1,\mathrm{ini}}(\bm{q}, \bm{k}) = 0.
  \label{initial_alpha1}
\end{align}
By inserting the expression \eqref{Psi_t} into the Shr\"odinger equation
$i \hbar \frac{\partial}{\partial t} \ket{\Psi(t)} = \hat{H} \ket{\Psi(t)}$,
we obtain
\begin{align}
   & \frac{\partial}{\partial t} \alpha_0(t, \bm{q})
  = -\frac{i}{\hbar} \int d^3q' \, d^3k' \, e^{i (\varepsilon_{\bm{q}} - \varepsilon_{\bm{q}'} - \omega_{\bm{k}'} ) t}
  \notag
  \\
   & \qquad \qquad \qquad \quad
  \times \bra{\bm{q}, \mathrm{vac}} \hat{H}_\mathrm{int} \ket{\bm{q}', \bm{k}'} \alpha_1(t, \bm{q}', \bm{k}'),
  \label{eq_alpha0}
  \\
   & \frac{\partial}{\partial t} \alpha_1(t, \bm{q}, \bm{k})
  = -\frac{i}{\hbar} \int d^3q' \, e^{i (\varepsilon_{\bm{q}} - \varepsilon_{\bm{q}'} + \omega_{\bm{k}} ) t}
  \notag
  \\
   & \qquad \qquad \qquad \qquad
  \times \bra{\bm{q}, \bm{k}} \hat{H}_\mathrm{int} \ket{\bm{q}', \mathrm{vac}} \alpha_0(t, \bm{q}').
  \label{eq_alpha1}
\end{align}

We expand the amplitudes up to the first order in $\hat{H}_\mathrm{int}$:
$\alpha_0(t, \bm{q}) = \alpha_0^{(0)}(t, \bm{q}) + \alpha_0^{(1)}(t, \bm{q})$ and
$\alpha_1(t, \bm{q}, \bm{k}) = \alpha_1^{(0)}(t, \bm{q}, \bm{k}) + \alpha_1^{(1)}(t, \bm{q}, \bm{k})$,
where $\alpha_0^{(j)}$ and $\alpha_1^{(j)}$ are the $j$th-order amplitudes.
The zeroth-order equations for Eqs.~\eqref{eq_alpha0} and \eqref{eq_alpha1} are respectively given by
\begin{align}
   & \frac{\partial}{\partial t} \alpha_0^{(0)}(t, \bm{q}) = 0,
  \\
   & \frac{\partial}{\partial t} \alpha_1^{(0)}(t, \bm{q}, \bm{k}) = 0.
\end{align}
Therefore the initial conditions \eqref{initial_alpha0} and \eqref{initial_alpha1} yield
\begin{align}
   & \alpha_0^{(0)}(t, \bm{q}) = \psi_\mathrm{ini}(\bm{q}),
  \label{alpha0_0}
  \\
   & \alpha_1^{(0)}(t, \bm{q}, \bm{k}) = 0.
  \label{alpha1_0}
\end{align}
With these zeroth-order solutions, the first-order equations for Eqs.~\eqref{eq_alpha0} and \eqref{eq_alpha1} are respectively given by
\begin{align}
   & \frac{\partial}{\partial t} \alpha_0^{(1)}(t, \bm{q})
  = 0,
  \label{1st_eq_alpha0}
  \\
   & \frac{\partial}{\partial t} \alpha_1^{(1)}(t, \bm{q}, \bm{k})
  = -\frac{i}{\hbar} \int d^3q' \, e^{i (\varepsilon_{\bm{q}} - \varepsilon_{\bm{q}'} + \omega_{\bm{k}} ) t}
  \notag
  \\
   & \qquad \qquad \qquad \qquad
  \times \bra{\bm{q}, \bm{k}} \hat{H}_\mathrm{int} \ket{\bm{q}', \mathrm{vac}} \psi_\mathrm{ini}(\bm{q}').
  \label{1st_eq_alpha1}
\end{align}
From Eq.~\eqref{1st_eq_alpha0}, we have
\begin{align}
  \alpha_0^{(1)}(t, \bm{q}) = 0.
  \label{alpha0_1}
\end{align}
We integrate Eq.~\eqref{1st_eq_alpha1} in time to obtain the amplitude after sufficiently long time as
\begin{align}
  \alpha_{1,\mathrm{fin}}^{(1)}(\bm{q}, \bm{k})
   & = -\frac{2 \pi i}{\hbar} \int d^3q' \, \delta (\varepsilon_{\bm{q}} - \varepsilon_{\bm{q}'} + \omega_{\bm{k}})
  \notag
  \\
   & \qquad \quad
  \times \bra{\bm{q}, \bm{k}} \hat{H}_\mathrm{int} \ket{\bm{q}', \mathrm{vac}} \psi_\mathrm{ini}(\bm{q}').
  \label{alpha1_1}
\end{align}

Putting Eqs.~\eqref{alpha0_0}, \eqref{alpha1_0}, \eqref{alpha0_1}, and \eqref{alpha1_1} together,
we obtain the state after the scattering as
\begin{align}
  \ket{\Psi_{\mathrm{fin}}'}
   & \propto \int d^3q \, \psi_\mathrm{ini}(\bm{q}) e^{-i\varepsilon_{\bm{q}} t} \ket{\bm{q}, \mathrm{vac}}
  \notag
  \\
   & \quad
  + \int d^3q \, d^3k \, \alpha_{1,\mathrm{fin}}^{(1)}(\bm{q}, \bm{k}) e^{-i(\varepsilon_{\bm{q}} + \omega_{\bm{k}}) t} \ket{\bm{q}, \bm{k}}.
\end{align}
The phase factors $e^{-i\varepsilon_{\bm{q}} t}$ and $e^{-i(\varepsilon_{\bm{q}} + \omega_{\bm{k}}) t}$ associated with the free propagation cancel out at the image plane when a lens system is employed.
Furthermore, by taking the normalization of the state into consideration, we arrive at
\begin{align}
  \ket{\Psi_{\mathrm{fin}}} = \sqrt{1 - a} \ket{\Psi_0} + \sqrt{a} \ket{\Psi_{\mathrm{sc}}}
\end{align}
with $a = A_1 / (A_0 + A_1)$, where
\begin{align}
  A_0 & = \int d^3q \, \bigl| \psi_\mathrm{ini}(\bm{q}) \bigr|^2 = 1,
  \\
  A_1 & = \int d^3q \, d^3k \, \bigl| \alpha_{1,\mathrm{fin}}^{(1)}(\bm{q}, \bm{k}) \bigr|^2,
\end{align}
and
\begin{align}
   \ket{\Psi_0} = 
   \int d^3q \, \psi_\mathrm{ini}(\bm{q}) \ket{\bm{q}, \mathrm{vac}}
   = \ket{\Psi_\mathrm{ini}},
\end{align}
and
\begin{align}
   & \ket{\Psi_{\mathrm{sc}}} = \int d^3q \, d^3k \, \psi_{\mathrm{sc}}(\bm{q}, \bm{k}) \ket{\bm{q}, \bm{k}},
  \label{Psi_scattered_0}
  \\
   & \psi_{\mathrm{sc}}(\bm{q}, \bm{k})
  = \frac{1}{\sqrt{A_1}} \alpha_{1,\mathrm{fin}}^{(1)}(\bm{q}, \bm{k}).
  \label{psi_sc_0}
\end{align}

\subsection{Connection between the scattered state and the luminescence spectrum}
\label{appendix:state_luminescence}

To obtain the expression \eqref{psi_sc} of the scattered-state amplitude $\psi_{\mathrm{sc}}(\bm{q}, \bm{k})$,
we consider the excitation probability (luminescence spectrum) $\Gamma(\bm{k}) = \int d^3q \, |\braket{\bm{q},\bm{k}}{\Psi_{\mathrm{sc}}}|^2$ of a photon with a wavevector $\bm{k}$ associated to this scattering event.
From Eqs.~\eqref{Psi_scattered_0} and\eqref{psi_sc_0}, it is given by
\begin{align}
  \Gamma(\bm{k}) = \int d^3q \, \bigl| \psi_{\mathrm{sc}}(\bm{q}, \bm{k}) \bigr|^2
  = \frac{1}{A_1} \int d^3q \, \bigl| \alpha_{1,\mathrm{fin}}^{(1)}(\bm{q}, \bm{k}) \bigr|^2.
  \label{excitation_prob}
\end{align}

To evaluate the right-hand side, we rewrite $\alpha_{1,\mathrm{fin}}^{(1)}(\bm{q}, \bm{k})$ in Eq.~\eqref{alpha1_1}.
As usual in coherent CL, we assume that the momentum change of the electron is sufficiently smaller than the mean momentum $\hbar q_0$ of an incident electron.
We can thus approximate
$\varepsilon_{\bm{q}} - \varepsilon_{\bm{q}'} \simeq \frac{\partial \varepsilon_{\bm{q}}}{\partial \bm{q}} \cdot (\bm{q} - \bm{q}') \simeq v_z (q_z - q'_z)$ with $v_z = \hbar q_0 / m$.
This approximation yields
\begin{align}
  \alpha_{1,\mathrm{fin}}^{(1)}(\bm{q}, \bm{k})
   & \simeq -\frac{2 \pi i}{\hbar v_z} \int d^3q' \, \delta \bigl( q_z - q'_z + ck / v_z \bigr)
  \notag
  \\
   & \qquad \quad \times \bra{\bm{q}, \bm{k}} \hat{H}_\mathrm{int} \ket{\bm{q}', \mathrm{vac}} \psi_\mathrm{ini}(\bm{q}').
\end{align}
Furthermore, by using the in-plane momentum-conservation condition given in Eq.~\eqref{in-plane-momentum-conservation}, we obtain
\begin{align}
  \alpha_{1,\mathrm{fin}}^{(1)}(\bm{q}, \bm{k})
  = -\frac{i}{(2 \pi)^2} \psi_\mathrm{ini}(\bm{q}_\perp + \bm{k}_\perp, q_z + ck / v_z)
  \, \tilde{\beta}(\bm{k}),
  \label{alpha1_1_fin_2}
\end{align}
where $\tilde{\beta}(\bm{k}) = (1/v_z) (2 \pi)^3 h(\bm{k}, -ck / v_z)$ is a function that depends only on $\bm{k}$.

By substituting Eq.~\eqref{alpha1_1_fin_2} into Eq.~\eqref{excitation_prob} and using the normalization condition for $\psi_\mathrm{ini}(\bm{q})$, we have
\begin{align}
  \Gamma(\bm{k})
   & = \frac{1}{(2 \pi)^4 A_1} \int d^3q \, \bigl| \psi_\mathrm{ini}(\bm{q}_\perp + \bm{k}_\perp, q_z + ck / v_z) \bigr|^2
  \, \bigl| \tilde{\beta}(\bm{k}) \bigr|^2
  \notag
  \\
   & = \frac{1}{(2 \pi)^4 A_1}\bigl| \tilde{\beta}(\bm{k}) \bigr|^2.
\end{align}
By inverting this relation, we obtain $\tilde{\beta}(\bm{k}) = i (2 \pi)^2 \sqrt{A_1 \Gamma(\bm{k})} e^{i\eta(\bm{k})}$,
where $\eta(\bm{k})$ is a phase that cannot be inferred from the luminescence spectrum $\Gamma(\bm{k})$ alone and may depend on $\bm{k}$.
From this relation and Eqs.~\eqref{psi_sc_0} and \eqref{alpha1_1_fin_2}, we arrive at the desired expression~\eqref{psi_sc} for $\psi_{\mathrm{sc}}(\bm{q}, \bm{k})$.
Since the luminescence spectrum $\Gamma(\bm{k})$ is experimentally accessible,
this relation enables us to reconstruct the scattered state $\ket{\Psi_{\mathrm{sc}}}$ of the electron--photon system from experimental data, up to the undetermined phase $\eta(\bm{k})$.
We note that since $\tilde{\beta}(\bm{k}) = (1/v_z) (2 \pi)^3 h(\bm{k}, -ck / v_z)$ and $h(\bm{k}, -ck / v_z)$ is invariant under rotations in the transverse plane (i.e., independent of the azimuthal angle $\phi$), $\Gamma(\bm{k})$ and $\eta(\bm{k})$ also inherit the same rotational symmetry.

\begin{widetext}
  \section{Calculation of the subsystem purity}
  \label{appendix:purity}

  In this appendix, we calculate the subsystem purity
  $\mathcal{P}_{\mathrm{sc}} = \mathrm{Tr}_{\mathrm{el}} \bigl[ ( \mathrm{Tr}_{\mathrm{ph}} \ketbra{\Psi_\mathrm{sc}}{\Psi_\mathrm{sc}} )^2 \bigr]$
  of the scattered state to derive Eq.~\eqref{purity}.

  The electron's reduced density matrix for the scattered state $\ket{\Psi_\mathrm{sc}}$ is given by
  \begin{align}
    \hat{\rho}_{\mathrm{el}}
     & = \mathrm{Tr}_{\mathrm{ph}} \ketbra{\Psi_\mathrm{sc}}{\Psi_\mathrm{sc}}
    \notag
    \\
     & = \int d^3 k \, \bra{\bm{k}}_{\mathrm{ph}} \bigl( \ketbra{\Psi_\mathrm{sc}}{\Psi_\mathrm{sc}} \bigr) \ket{\bm{k}}_{\mathrm{ph}}
    \notag
    \\
     & = \int d^3 \, q d^3 q' \, d^3 k \, \psi_\mathrm{sc}^*(\bm{q}', \bm{k}) \, \psi_\mathrm{sc}(\bm{q}, \bm{k}) \, \ket{\bm{q}}_{\mathrm{el}} \bra{\bm{q}'}_{\mathrm{el}}.
    \notag
  \end{align}
  We then calculate $\mathcal{P}_{\mathrm{sc}}$ as
  \begin{align}
    \mathcal{P}_{\mathrm{sc}}
     & = \mathrm{Tr}_{\mathrm{el}} \bigl[ \hat{\rho}_{\mathrm{el}}^2 \bigr]
    \notag
    \\
     & = \int d^3 q \, \bra{\bm{q}}_{\mathrm{el}} \hat{\rho}_{\mathrm{el}}^2 \ket{\bm{q}}_{\mathrm{el}}
    \notag
    \\
     & = \int d^3 q \, d^3 q' \, d^3 k \, d^3 k' \, \psi_\mathrm{sc}^*(\bm{q}', \bm{k}) \, \psi_\mathrm{sc}(\bm{q}, \bm{k}) \, \psi_\mathrm{sc}^*(\bm{q}', \bm{k}') \, \psi_\mathrm{sc}(\bm{q}, \bm{k}')
    \notag
    \\
     & = \int d^3 k \, d^3 k' \, \Gamma(\bm{k}) \, \Gamma(\bm{k}')
    \biggl[ \int d^3 q \, \psi_\mathrm{ini}(\bm{q}_\perp + \bm{k}_\perp, q_z + \frac{ck}{v_z}) \, \psi_\mathrm{ini}(\bm{q}_\perp + \bm{k}'_\perp, q_z + \frac{ck'}{v_z}) \biggr]^2.
    \label{purity_0}
  \end{align}
  We have used the fact that $\psi_\mathrm{ini}(\bm{q})$ is a real-valued function.
  Using Eqs.~\eqref{initial_psi_xy} and \eqref{initial_psi_z} for $\psi_\mathrm{ini}(\bm{q})$,
  we can calculate the integral in the square bracket of the above equation as
  \begin{align}
     & \int d^3 q \, \psi_\mathrm{ini}(\bm{q}_\perp + \bm{k}_\perp, q_z + ck/v_z) \, \psi_\mathrm{ini}(\bm{q}_\perp + \bm{k}'_\perp, q_z + ck'/v_z)
    =\exp[ -\frac{|\bm{k}_\perp - \bm{k}'_\perp|^2}{8 (\sigma_{q,\perp}^\mathrm{el})^2}
      - \frac{c^2 (k - k')^2}{8 v_z^2 (\sigma_{q,\parallel}^\mathrm{el})^2} ].
  \end{align}
  Substituting this result into Eq.~\eqref{purity_0}, we obtain Eq.~\eqref{purity}.

  \section{Calculation of the joint probability distributions}

  \subsection{Joint spatial distribution $P(x_\mathrm{el},x_\mathrm{ph})$}
  \label{appendix:dist_x_x}

  We here derive Eq.~\eqref{dist_x_x} for the joint probability distribution $P(x_\mathrm{el},x_\mathrm{ph})$ that the electron is at $x_\mathrm{el}$ and the photon is at $x_\mathrm{ph}$ in the $x$ direction.

  The joint distribution $P(x_\mathrm{el}, x_\mathrm{ph})$ is defined by
  \begin{align}
    P(x_\mathrm{el}, x_\mathrm{ph})
    = \int dy_{\mathrm{el}} dz_{\mathrm{el}} \, dy_{\mathrm{ph}} dz_{\mathrm{ph}}
    \, \bigl| \braket{\bm{r}_{\mathrm{el}}, \bm{r}_{\mathrm{ph}}}{\Psi_\mathrm{sc}} \bigr|^2,
    \label{dist_x_x_def}
  \end{align}
  where $\bm{r}_\mathrm{el} = (x_\mathrm{el}, y_\mathrm{el}, z_\mathrm{el})$ and $\bm{r}_\mathrm{ph} = (x_\mathrm{ph}, y_\mathrm{ph}, z_\mathrm{ph})$.

  Using Eqs.~\eqref{Psi_scattered} and \eqref{psi_sc} and $\braket{\bm{r}_{\mathrm{el}}, \bm{r}_{\mathrm{ph}}}{\bm{q}, \bm{k}} = e^{i\bm{q}\cdot\bm{r}_{\mathrm{el}} + i\bm{k}\cdot\bm{r}_{\mathrm{ph}}} / (2\pi)^3$,
  we can calculate the inner product in the above equation as
  \begin{align}
     & \braket{\bm{r}_{\mathrm{el}}, \bm{r}_{\mathrm{ph}}}{\Psi_\mathrm{sc}}
    = \int \frac{d^3q \, d^3k}{(2\pi)^3} \psi_\mathrm{ini}(\bm{q}_\perp + \bm{k}_\perp, q_z + \frac{ck}{v_z})
    \sqrt{\Gamma(\bm{k})} e^{i\eta(\bm{k})} e^{i\bm{q}\cdot\bm{r}_{\mathrm{el}} + i\bm{k}\cdot\bm{r}_{\mathrm{ph}}}
    \notag
    \\
     & = \biggl(\frac{2}{\pi}\biggr)^{3/4} \sigma_{q,\perp}^\mathrm{el} \bigl( \sigma_{q,\parallel}^\mathrm{el} \bigr)^{1/2}
    e^{-(\sigma_{q,\perp}^\mathrm{el})^2 |\bm{r}_\perp^\mathrm{el}|^2 - (\sigma_{q,\parallel}^\mathrm{el})^2 z_\mathrm{el}^2 + iq_0 z_\mathrm{el}}
    \int \frac{d^3k}{(2\pi)^{3/2}} \sqrt{\Gamma(\bm{k})} e^{-i\bm{k}_\perp \cdot (\bm{r}_\perp^\mathrm{el} - \bm{r}_\perp^\mathrm{ph})}
    e^{-i\frac{ck}{v_z} z_{\mathrm{el}} + ik_z z_{\mathrm{ph}}} e^{i\eta(\bm{k})},
  \end{align}
  where $\bm{r}_\perp^\mathrm{el} = (x_\mathrm{el}, y_\mathrm{el})$ and $\bm{r}_\perp^\mathrm{ph} = (x_\mathrm{ph}, y_\mathrm{ph})$.
  In the second equality, we have performed the integration over $\bm{q}$ with Eqs.~\eqref{initial_psi_xy} and \eqref{initial_psi_z} for $\psi_\mathrm{ini}(\bm{q})$.
  Therefore, Eq.~\eqref{dist_x_x_def} yields
  \begin{align}
    P(x_\mathrm{el}, x_\mathrm{ph})
     & = \int dy_{\mathrm{el}} \, dz_{\mathrm{el}} \, dy_{\mathrm{ph}} \, dz_{\mathrm{ph}}
    \biggl(\frac{2}{\pi}\biggr)^{3/2} \bigl( \sigma_{q,\perp}^\mathrm{el} \bigr)^2 \sigma_{q,\parallel}^\mathrm{el}
    e^{-2(\sigma_{q,\perp}^\mathrm{el})^2 |\bm{r}_\perp^\mathrm{el}|^2 - 2(\sigma_{q,\parallel}^\mathrm{el})^2 z_\mathrm{el}^2}
    \notag
    \\
     & \qquad \qquad \qquad
    \times \int \frac{d^3k \, d^3k'}{(2\pi)^3} \sqrt{\Gamma(\bm{k}) \Gamma(\bm{k}')} e^{-i(\bm{k}_\perp - \bm{k}_\perp') \cdot (\bm{r}_\perp^\mathrm{el} - \bm{r}_\perp^\mathrm{ph})}
    e^{-i\frac{c(k - k')}{v_z} z_{\mathrm{el}} + i(k_z - k_z') z_{\mathrm{ph}}} e^{i[\eta(\bm{k})-\eta(\bm{k}')]}
    \notag
    \\
     & = \frac{1}{2\pi} \biggl(\frac{2}{\pi}\biggr)^{3/2} \bigl( \sigma_{q,\perp}^\mathrm{el} \bigr)^2 \sigma_{q,\parallel}^\mathrm{el}
    \int dy_{\mathrm{el}} \, dz_{\mathrm{el}} \, d^3k \, d^3k'
    e^{-2(\sigma_{q,\perp}^\mathrm{el})^2 |\bm{r}_\perp^\mathrm{el}|^2 - 2(\sigma_{q,\parallel}^\mathrm{el})^2 z_\mathrm{el}^2}
    \sqrt{\Gamma(\bm{k}) \Gamma(\bm{k}')}
    \notag
    \\
     & \qquad \qquad \qquad \qquad \qquad \qquad
    \times \delta(k_y - k_y') \delta(k_z - k_z')
    e^{-i(\bm{k}_\perp - \bm{k}_\perp') \cdot \bm{r}_\perp^\mathrm{el}} e^{i(k_x - k_x') x_\mathrm{ph}}
    e^{-i\frac{c(k - k')}{v_z} z_{\mathrm{el}}} e^{i[\eta(\bm{k})-\eta(\bm{k}')]}
    \notag
    \\
     & = \frac{1}{2\pi} \biggl(\frac{2}{\pi}\biggr)^{3/2} \bigl( \sigma_{q,\perp}^\mathrm{el} \bigr)^2 \sigma_{q,\parallel}^\mathrm{el}
    \int dy_{\mathrm{el}} \, dz_{\mathrm{el}} \, d^3k \, dk_x'
    e^{-2(\sigma_{q,\perp}^\mathrm{el})^2 |\bm{r}_\perp^\mathrm{el}|^2 - 2(\sigma_{q,\parallel}^\mathrm{el})^2 z_\mathrm{el}^2}
    \sqrt{\Gamma(\bm{k}) \Gamma(k_x', k_y, k_z)}
    \notag
    \\
     & \qquad \qquad \qquad \qquad \qquad \qquad
    \times e^{-i(k_x - k_x') (x_\mathrm{el} - x_\mathrm{ph})}
    e^{-i\frac{c}{v_z} (k - \sqrt{k_x^{\prime 2} + k_y^2 + k_z^2}) z_{\mathrm{el}}} e^{i[\eta(\bm{k})-\eta(k_x', k_y, k_z)]}.
  \end{align}
  In the second equality, we have used $\int dy \, e^{i(k_y - k_y')y}= 2\pi \delta(k_y - k_y')$.
  Furthermore by performing the integration over $y_{\mathrm{el}}$ and $z_{\mathrm{el}}$, we obtain Eq.~\eqref{dist_x_x}.

  \subsection{Joint wavevector distribution $P(q_x, k_x)$}
  \label{appendix:dist_qx_kx}

  We here derive Eq.~\eqref{dist_qx_kx} for the joint probability distribution $P(q_x, k_x)$ that the electron's wavevector is $q_x$ and the photon's wavevector $k_x$ in the $x$ direction.

  The joint distribution $P(q_x, k_x)$ is defined by
  \begin{align}
    P(q_x, k_x)
    = \int dq_y \, dq_z \, dk_y \, dk_z \, \bigl| \braket{\bm{q}, \bm{k}}{\Psi_\mathrm{sc}} \bigr|^2,
    \label{dist_qx_kx_def}
  \end{align}
  where $\bm{q} = (q_x, q_y, q_z)$ and $\bm{k} = (k_x, k_y, k_z)$.
  From Eqs.~\eqref{Psi_scattered} and \eqref{psi_sc}, we can calculate $P(\bm{q}, \bm{k})$ as
  \begin{align}
    P(\bm{q}, \bm{k})
    = \bigl| \psi_\mathrm{sc}(\bm{q}, \bm{k}) \bigr|^2
    = \Bigl| \psi_\mathrm{ini}(\bm{q}_\perp + \bm{k}_\perp, q_z + \frac{ck}{v_z}) \Bigr|^2 \, \Gamma(\bm{k}).
  \end{align}
  Substituting this into Eq.~\eqref{dist_qx_kx_def}
  and performing the integration over $q_y$ and $q_z$ with the normalization conditions for $\psi_\mathrm{ini}^{(y)}$ and $\psi_\mathrm{ini}^{(z)}$,
  we obtain Eq.~\eqref{dist_qx_kx}.

  \section{Calculation of the uncertainties}
  \label{appendix:uncertainty}

  \subsection{Relative position uncertainty $\expval{[\Delta (x_\mathrm{el} - x_\mathrm{ph})]^2}$}
  \label{appendix:Delta_x_x}

  We here derive Eqs.~\eqref{Delta_x_x} and \eqref{Delta_x_x_model} for the relative position uncertainty $\expval{[\Delta (x_\mathrm{el} - x_\mathrm{ph})]^2}$.

  We first show that the average of $x_\mathrm{el} - x_\mathrm{ph}$ is zero.
  To this end, we note the relations $\Gamma(-k_x, k_y, k_z) = \Gamma(k_x, k_y, k_z)$ and $\eta(-k_x, k_y, k_z) = \eta(k_x, k_y, k_z)$,
  which follow from the in-plane rotational symmetry for $\Gamma(\bm{k})$ and $\eta(\bm{k})$.
  Equation~\eqref{dist_x_x} with these relations leads to $P(-x_{\mathrm{el}}, -x_{\mathrm{ph}}) = P(x_{\mathrm{el}}, x_{\mathrm{ph}})$.
  Therefore, $\expval{x_\mathrm{el} - x_\mathrm{ph}} = \int dx_{\mathrm{el}} \, dx_{\mathrm{ph}} \, P(x_{\mathrm{el}}, x_{\mathrm{ph}}) \, (x_\mathrm{el} - x_\mathrm{ph}) = 0$.

  We then calculate the uncertainty $\expval{[\Delta (x_\mathrm{el} - x_\mathrm{ph})]^2} = \expval{(x_\mathrm{el} - x_\mathrm{ph})^2}$.
  From Eq.~\eqref{dist_x_x}, we calculate it as
  \begin{align}
     & \expval{[\Delta (x_\mathrm{el} - x_\mathrm{ph})]^2}
    = \int dx_{\mathrm{el}} \, dx_{\mathrm{ph}} \, P(x_{\mathrm{el}}, x_{\mathrm{ph}}) \, (x_\mathrm{el} - x_\mathrm{ph})^2
    \notag
    \\
     & = \int dx_{\mathrm{el}} \, dx_{\mathrm{ph}} \frac{\sigma_{q,\perp}^\mathrm{el}}{\sqrt{2\pi^3}}
    e^{-2(\sigma_{q,\perp}^\mathrm{el})^2 x_\mathrm{el}^2}
    \int d^3k \, dk_x'
    \sqrt{\Gamma(\bm{k}) \Gamma(\bm{k}')}
    \biggl( -\frac{\partial^2}{\partial k_x^{\prime 2}} e^{-i(k_x - k_x') (x_\mathrm{el} - x_\mathrm{ph})} \biggr)
    e^{-c^2(k - k')^2 / [8v_z^2 (\sigma_{q,\parallel}^\mathrm{el})^2]} e^{i[\eta(\bm{k})-\eta(\bm{k}')]}
    \notag
    \\
     & = \int dx_{\mathrm{el}} \, dx_{\mathrm{ph}} \, d^3k \, dk_x' \frac{\sigma_{q,\perp}^\mathrm{el}}{\sqrt{2\pi^3}}
    e^{-2(\sigma_{q,\perp}^\mathrm{el})^2 x_\mathrm{el}^2}
    e^{-i(k_x - k_x') (x_\mathrm{el} - x_\mathrm{ph})}
    \biggl( -\frac{\partial^2}{\partial k_x^{\prime 2}} \sqrt{\Gamma(\bm{k}) \Gamma(\bm{k}')}
    e^{-c^2(k - k')^2 / [8v_z^2 (\sigma_{q,\parallel}^\mathrm{el})^2]} e^{i[\eta(\bm{k})-\eta(\bm{k}')]} \biggr)
    \notag
    \\
     & = \int dx_{\mathrm{el}} \, d^3k \, dk_x' \frac{\sigma_{q,\perp}^\mathrm{el}}{\sqrt{2\pi^3}} 2\pi \delta(k_x - k_x')
    e^{-2(\sigma_{q,\perp}^\mathrm{el})^2 x_\mathrm{el}^2}
    \biggl( -\frac{\partial^2}{\partial k_x^{\prime 2}} \sqrt{\Gamma(\bm{k}) \Gamma(\bm{k}')}
    e^{-c^2(k - k')^2 / [8v_z^2 (\sigma_{q,\parallel}^\mathrm{el})^2]} e^{i[\eta(\bm{k})-\eta(\bm{k}')]} \biggr)
    \notag
    \\
     & = \int d^3k \, dk_x' \delta(k_x - k_x')
    \biggl( -\frac{\partial^2}{\partial k_x^{\prime 2}} \sqrt{\Gamma(\bm{k}) \Gamma(\bm{k}')}
    e^{-c^2(k - k')^2 / [8v_z^2 (\sigma_{q,\parallel}^\mathrm{el})^2]} e^{i[\eta(\bm{k})-\eta(\bm{k}')]} \biggr)
    \notag
    \\
     & = \int d^3k \left\{
    \frac{1}{4\Gamma(\bm{k})} \biggl( \frac{\partial \Gamma(\bm{k})}{\partial k_x} \biggr)^2
    - \frac{1}{2} \frac{\partial^2 \Gamma(\bm{k})}{\partial k_x^2}
    + \frac{c^2 k_x^2}{4 v_z^2 (\sigma_{q,\parallel}^\mathrm{el})^2 k^2} \Gamma(\bm{k})
    + \Gamma(\bm{k}) \biggl( \frac{\partial \eta(\bm{k})}{\partial k_x} \biggr)^2
    \right\}.
    \label{D_x_x_sm_1}
  \end{align}
  Furthermore, from the rotational symmetry in the transverse plane, we have
  \begin{align}
    \expval{[\Delta (x_\mathrm{el} - x_\mathrm{ph})]^2} = \expval{[\Delta (y_\mathrm{el} - y_\mathrm{ph})]^2} = \frac{1}{2} \bigl\{\expval{[\Delta (x_\mathrm{el} - x_\mathrm{ph})]^2} + \expval{[\Delta (y_\mathrm{el} - y_\mathrm{ph})]^2} \bigr\}.
  \end{align}
  Since $\expval{[\Delta (y_\mathrm{el} - y_\mathrm{ph})]^2}$ is obtained from Eq.~\eqref{D_x_x_sm_1} by replacing $k_x$ with $k_y$,
  the desired Eq.~\eqref{Delta_x_x} follows from these relations.

  Next, we assume the model for $\Gamma(\bm{k})$ [Eqs.~\eqref{Gamma_model}--\eqref{f_model}] and further evaluate $\expval{[\Delta (x_\mathrm{el} - x_\mathrm{ph})]^2}$ in order to derive Eq.~\eqref{Delta_x_x_model}.
  To this end, we use the spherical coordinates ($k_x = k \sin\theta \cos\phi$, $k_y = k \sin\theta \sin\phi$, $k_z = k \cos\theta$ with the polar angle $\theta$ and azimuthal angle $\phi$).
  We calculate the derivatives in Eq.~\eqref{Delta_x_x} by noting $\partial \Gamma(\bm{k}) / \partial \phi = 0$ due to the rotational symmetry in the transverse plane.
  We then perform the integration in Eq.~\eqref{Delta_x_x} to obtain
  \begin{align}
     & \frac{1}{8} \int d^3k \left\{ \frac{1}{\Gamma(\bm{k})} \left[ \biggl(\frac{\partial \Gamma(\bm{k})}{\partial k_x} \biggr)^2 + \biggl(\frac{\partial \Gamma(\bm{k})}{\partial k_y} \biggr)^2 \right] - 2 \biggl( \frac{\partial^2 \Gamma(\bm{k})}{\partial k_x^2} + \frac{\partial^2 \Gamma(\bm{k})}{\partial k_y^2} \biggr)
    \right\}
    \notag
    \\
     & = \frac{15 \mathcal{N}_g}{64 \pi (\sigma_{k}^\mathrm{ph})^4} \int_0^\infty dk \int_0^\pi d\theta \int_0^{2\pi} d\phi \, \sin\theta \, \cos^2\theta \, e^{-(k -k_c)^2/[2 (\sigma_{k}^\mathrm{ph})^2]}
    \biggl(  - 4 (\sigma_{k}^\mathrm{ph})^4
    + 8 (\sigma_{k}^\mathrm{ph})^2 \bigl[ k^2 - k_c k + 4 (\sigma_{k}^\mathrm{ph})^2 \bigr] \sin^2\theta
    \notag
    \\
     & \qquad \qquad \qquad \qquad \qquad \qquad \qquad \qquad \qquad \qquad \qquad
    - \Bigl\{ k^4 - 2 k_c k^3 + [k_c^2 + 8 (\sigma_{k}^\mathrm{ph})^2] k^2 - 10 k_c (\sigma_{k}^\mathrm{ph})^2 k + 32 (\sigma_{k}^\mathrm{ph})^4 \Bigr\} \sin^4\theta \biggr)
    \notag
    \\
     & = \frac{\mathcal{N}_g \sqrt{2 \pi}}{56}
    \biggl( 19 \sigma_{k}^\mathrm{ph} + \frac{2 k_c^2}{\sigma_{k}^\mathrm{ph}} \biggr) \biggl[ \erf\biggl(\frac{k_c}{\sqrt{2} \sigma_{k}^\mathrm{ph}} \biggr) + 1 \biggr]
    + \frac{\mathcal{N}_g}{14} k_c \exp \biggl( -\frac{k_c^2}{2 (\sigma_{k}^\mathrm{ph})^2}\biggr),
  \end{align}
  and
  \begin{align}
    \frac{1}{8} \int d^3k\frac{c^2 k_\perp^2 \Gamma(\bm{k})}{v_z^2 (\sigma_{q,\parallel}^\mathrm{el})^2 k^2}
    =\frac{c^2}{8 v_z^2 (\sigma_{q,\parallel}^\mathrm{el})^2} \int_0^\infty dk \int_0^\pi d\theta \int_0^{2\pi} d\phi \, k^2 \sin^3\theta \, \Gamma(\bm{k})
    = \frac{c^2}{14 v_z^2 (\sigma_{q,\parallel}^\mathrm{el})^2}.
  \end{align}
  We thus obtain Eq.~\eqref{Delta_x_x_model}.

  \subsection{Total wavevector uncertainty $\expval{[\Delta (q_x + k_x)]^2}$}
  \label{appendix:Delta_qx_kx}

  We here derive Eq.~\eqref{Delta_qx_kx}, $\expval{[\Delta (q_x + k_x)]^2} = \bigl( \sigma_{q,\perp}^\mathrm{el} \bigr)^2$,
  without assuming a model for $\Gamma(\bm{k})$.

  We first show that the average of $q_x + k_x$ is zero.
  Using Eqs.~\eqref{dist_qx_kx} and \eqref{initial_psi_xy} and noting $\big| \psi^{(x)}_\mathrm{ini}(-q_x) |^2 = | \psi^{(x)}_\mathrm{ini}(q_x) |^2$,
  we calculate $\expval{q_x + k_x}$ as
  \begin{align}
    \expval{q_x + k_x}
     & = \int dq_x \, dk_x \, P(q_x, k_x) \, (q_x + k_x)
    \notag
    \\
     & = \int dk_x \, dk_y \, dk_z \, \Gamma(\bm{k})
    \int dq_x \, \big| \psi^{(x)}_\mathrm{ini}(q_x + k_x) \big|^2 \, (q_x + k_x)
    \notag
    \\
     & = 0.
     \label{q_x_k_x_avg}
  \end{align}
  In deriving this result, we do not assume $q_x = -k_x$. Rather, $q_x + k_x$ is distributed according to the transverse momentum distribution of the incident electron because the scattered-state amplitude contains $\psi^{(x)}_\mathrm{ini}$ evaluated at $q_{\mathrm{in},x} = q_x + k_x$. 
  Since the incident transverse momentum distribution is centered at zero and is symmetric, its average vanishes, yielding Eq.~\eqref{q_x_k_x_avg}. 
  For a finite transverse coherence length, $q_x + k_x$ is therefore not identically zero but has a finite width determined by $\sigma_{q,\perp}^\mathrm{el}$, as shown in the following.

  We then investigate the uncertainty
  $\expval{[\Delta (q_x + k_x)]^2} = \expval{(q_x + k_x)^2}$.
  Again, using Eqs.~\eqref{dist_qx_kx} and \eqref{initial_psi_xy}, we obtain the desired result:
  \begin{align}
    \expval{[\Delta (q_x + k_x)]^2}
     & = \int dq_x \, dk_x \, P(q_x, k_x) \, (q_x + k_x)^2
    \notag
    \\
     & = \int dk_x \, dk_y \, dk_z \, \Gamma(\bm{k})
    \int dq_x \, \big| \psi^{(x)}_\mathrm{ini}(q_x + k_x) \big|^2 \, (q_x + k_x)^2
    \notag
    \\
     & = \bigl( \sigma_{q,\perp}^\mathrm{el} \bigr)^2.
  \end{align}
\end{widetext}
We emphasize that we do not assume that the variance of the scattered electron's transverse wavevector is equal to $(\sigma_{q,\perp}^\mathrm{el})^2$. Rather, we derive that the variance of the total transverse wavevector component $q_x + k_x$ in the scattered electron--photon state is equal to the corresponding variance of the incident electron, $(\sigma_{q,\perp}^\mathrm{el})^2$. As explained below Eq.~\eqref{Delta_qx_kx}, this result follows from conservation of the total transverse momentum of the electron--photon system.

\begin{figure}[htb]
  \includegraphics[width=\linewidth]{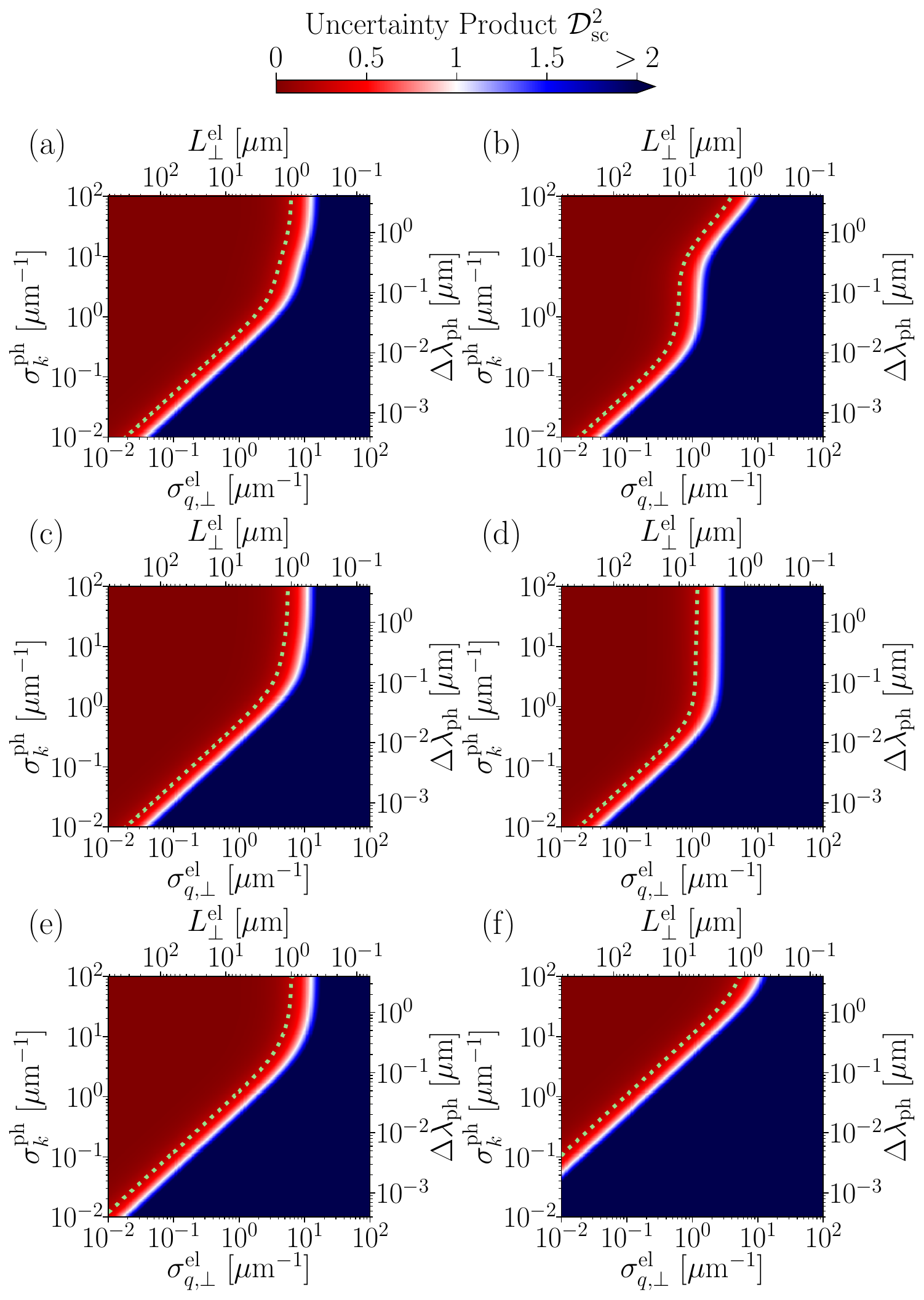}
  \caption{\label{fig:uncertainty_phase}
  Influence of the phase $\eta(\bm{k})$ on the uncertainty product $\mathcal{D}_{\mathrm{sc}}^2$.
  Panels (a) and (b) correspond to the case where $\eta(\bm{k})$ depends only on $\theta$ and the additional term $D_\eta$ is given by Eq.~\eqref{D_eta_1} with $\xi_1 = 1$ for (a) and $\xi_1 = 100$ for (b).
  Panels (c) and (d) correspond to the case of $\eta_{21}(k) = \sqrt{\xi_2} k / k_c$ with $\xi_2 = 1$ for (c) and $\xi_2 = 100$ for (d).
  Panels (e) and (f) correspond to the case of $\eta_{22}(k) = \sqrt{\xi_2} k / \sigma_{k}^\mathrm{ph}$ with $\xi_2 = 1$ for (e) and $\xi_2 = 100$ for (f).
  The other parameter values are the same as those in Fig.~\ref{fig:entanglement_1}.
  The dotted curves represent the EPR-correlation threshold $\mathcal{D}_{\mathrm{sc}}^2 = 1/4$.}
\end{figure}

\section{Influence of the phase $\eta(\texorpdfstring{\bm{k}}{k})$}
\label{appendix:uncertainty_phase}

In Sec.~\ref{sec:EPR}, to discuss the entanglement intrinsic to the scattered state $\ket{\Psi_\mathrm{sc}}$,
we investigate the uncertainty product $\mathcal{D}_{\mathrm{sc}}^2$, focusing on the ideal case in which the phase-dependent term $D_\eta$ vanishes in the relative position uncertainty $\expval{[\Delta (x_\mathrm{el} - x_\mathrm{ph})]^2}$.
However, in practical experiments, the condition $D_\eta = 0$ may not be perfectly achieved, and thus $\mathcal{D}_{\mathrm{sc}}^2$ may appear to increase.
In this appendix, we therefore investigate the influence of the phase $\eta(\bm{k})$ on $\mathcal{D}_{\mathrm{sc}}^2$ by assuming several models for $\eta(\bm{k})$.
We note that, while $\expval{[\Delta (x_\mathrm{el} - x_\mathrm{ph})]^2}$ contains the phase-dependent term $D_\eta$, as seen in Eqs.~\eqref{Delta_x_x} and \eqref{Delta_x_x_model},
the total wavevector uncertainty $\expval{[\Delta (q_x + k_x)]^2}$ is independent of the phase, as seen in Eq.~\eqref{Delta_qx_kx}.

We first consider the case where the phase $\eta(\bm{k})$ depends only on the polar angle $\theta$, i.e., $\eta(\bm{k}) = \eta_1(\theta)$.
In this case, $D_\eta$ reduces to
\begin{align}
  D_{\eta_1} & = \xi_1 \sqrt{\frac{\pi}{2}} \mathcal{N}_g \sigma_{k}^\mathrm{ph}
  \biggl[ \erf\biggl(\frac{k_c}{\sqrt{2} \sigma_{k}^\mathrm{ph}} \biggr) + 1 \biggr],
  \label{D_eta_1}
\end{align}
where the dimensionless factor $\xi_1$ is defined as
\begin{align}
  \xi_1 & = \pi \int_0^\pi d\theta \, \sin\theta \, \cos^2\theta \, f(\theta) \left( \frac{\partial \eta_1(\theta)}{\partial \theta} \right)^2,
\end{align}
with $f(\theta)$ given by Eq.~\eqref{f_model}.
For example, $\xi_1 = 3/14$ when $\eta_1(\theta) = \theta$.

Figures~\ref{fig:uncertainty_phase}(a) and (b) show color plots of $\mathcal{D}_{\mathrm{sc}}^2$ for this case with $\xi_1 = 1$ and $\xi_1 = 100$, respectively.
Compared with Fig.~\ref{fig:entanglement_1}(c), the result for $\xi_1 = 1$ remains nearly unchanged, whereas a clear deviation appears for $\xi_1 = 100$ in the large-$\sigma_{k}^\mathrm{ph}$ regime.
This behavior can be understood from the relation between $D_{\eta_1}$ and the term
\begin{align}
  D_0 = \frac{19 \sqrt{2 \pi}}{56} \mathcal{N}_g \sigma_{k}^\mathrm{ph} \biggl[ \erf\biggl(\frac{k_c}{\sqrt{2} \sigma_{k}^\mathrm{ph}} \biggr) + 1 \biggr],
\end{align}
which appears in $\expval{[\Delta (x_\mathrm{el} - x_\mathrm{ph})]^2}$ [Eq.~\eqref{Delta_x_x_model}].
Since $D_{\eta_1} \propto \xi_1 D_0$, its relative importance is controlled by $\xi_1$.
For $\xi_1 \lesssim 1$, one has $D_{\eta_1} \lesssim D_0$, so its contribution to $\expval{[\Delta (x_\mathrm{el} - x_\mathrm{ph})]^2}$ is minor and $\mathcal{D}_{\mathrm{sc}}^2$ remains essentially unchanged.
In contrast, for $\xi_1 \gg 1$, $D_{\eta_1}$ dominates at large $\sigma_{k}^\mathrm{ph}$, leading to $\mathcal{D}_{\mathrm{sc}}^2 \simeq D_{\eta_1} (\sigma_{q,\perp}^\mathrm{el})^2$.
Furthermore, around $\sigma_{k}^\mathrm{ph} \approx k_c$, the dependence of $D_{\eta_1}$ on $\sigma_{k}^\mathrm{ph}$ is weak, so the contour $\mathcal{D}_{\mathrm{sc}}^2 = 1/4$ becomes nearly vertical in the $\sigma_{q,\perp}^\mathrm{el}$--$\sigma_{k}^\mathrm{ph}$ plane.
For $\sigma_{k}^\mathrm{ph} \gg k_c$, $D_{\eta_1} \propto 1 / (\sigma_{k}^\mathrm{ph})^2$, which implies $\sigma_{q,\perp}^\mathrm{el} \propto \sigma_{k}^\mathrm{ph}$ along the $\mathcal{D}_{\mathrm{sc}}^2 = 1/4$ contour for sufficiently large $\sigma_{k}^\mathrm{ph}$.

We next consider the case where the phase $\eta(\bm{k})$ depends only on $k = |\bm{k}|$, i.e., $\eta(\bm{k}) = \eta_2(k)$.
In this case, $D_\eta$ reduces to
\begin{align}
  D_{\eta_2} & = \frac{2}{7} \int_0^\infty dk \, k^2 g(k) \left( \frac{\partial \eta_2(k)}{\partial k} \right)^2,
  \label{D_eta_2}
\end{align}
where $g(k)$ is given by Eq.~\eqref{g_model}.
Since $\eta_2(k)$ is dimensionless, it may be a function of $k / k_c$ and $k / \sigma_{k}^\mathrm{ph}$.
As minimal models, we consider $\eta_{21}(k) = \sqrt{\xi_2} k / k_c$ and $\eta_{22}(k) = \sqrt{\xi_2} k / \sigma_{k}^\mathrm{ph}$ with a dimensionless constant $\xi_2$.
These models yield $D_{\eta_{21}} = 2 \xi_2 / (7 k_c^2)$ and $D_{\eta_{22}} = 2 \xi_2 / [7 (\sigma_{k}^\mathrm{ph})^2]$, respectively.

Figures~\ref{fig:uncertainty_phase}(c) and (d) show color plots of $\mathcal{D}_{\mathrm{sc}}^2$ for $\eta_{21}(k) = \sqrt{\xi_2} k / k_c$ with $\xi_2 = 1$ and $\xi_2 = 100$, respectively.
Compared with Fig.~\ref{fig:entanglement_1}(c), the result for $\xi_2 = 1$ is nearly unchanged, whereas for $\xi_2 = 100$ the vertical segment of the $\mathcal{D}_{\mathrm{sc}}^2 = 1/4$ contour becomes longer and shifts to smaller $\sigma_{k}^\mathrm{ph}$.
This behavior can be understood from the fact that $D_{\eta_{21}} = 2\xi_2/(7 k_c^2)$ is independent of $\sigma_{k}^\mathrm{ph}$, whereas the other contributions to $\expval{[\Delta (x_\mathrm{el} - x_\mathrm{ph})]^2}$ either decrease or remain unchanged as $\sigma_{k}^\mathrm{ph}$ increases.
As a result, for large $\sigma_{k}^\mathrm{ph}$ and $\xi \gg 1$, $D_{\eta_{21}}$ becomes the dominant contribution.
Moreover, a larger $\xi_2$ accelerates this crossover to the $D_{\eta_{21}}$-dominated regime.
In this regime, one finds $\mathcal{D}_{\mathrm{sc}}^2 \propto \xi_2 (\sigma_{q,\perp}^\mathrm{el})^2$, so that the contour $\mathcal{D}_{\mathrm{sc}}^2 = 1/4$ becomes nearly vertical in the $\sigma_{q,\perp}^\mathrm{el}$--$\sigma_{k}^\mathrm{ph}$ plane, as seen in Fig.~7(d).

Figures~\ref{fig:uncertainty_phase}(e) and (f) show color plots of $\mathcal{D}_{\mathrm{sc}}^2$ for $\eta_{22}(k) = \sqrt{\xi_2} k / \sigma_{k}^\mathrm{ph}$ with $\xi_2 = 1$ and $\xi_2 = 100$, respectively.
Compared with Fig.~\ref{fig:entanglement_1}(c), the $\mathcal{D}_{\mathrm{sc}}^2 = 1/4$ contour shifts upward as $\xi_2$ increases.
This behavior follows directly from the form of $D_{\eta_{22}} = 2\xi_2/[7 (\sigma_{k}^\mathrm{ph})^2]$, which enhances the contribution [$\propto 1 / (\sigma_{k}^\mathrm{ph})^2$] in $\expval{[\Delta (x_\mathrm{el} - x_\mathrm{ph})]^2}$ at small $\sigma_{k}^\mathrm{ph}$ for larger $\xi_2$.

\bibliography{refs.bib}

\end{document}